\documentclass[journal,12pt,one column]{IEEEtran}
\usepackage{amsfonts}
\usepackage{amsmath}    
\usepackage{graphicx}   
\usepackage{graphics}   
\usepackage{verbatim}   
\usepackage{color}      
\usepackage{subfigure}  
\usepackage{array}
\usepackage{amssymb}
\usepackage{sidecap}
\usepackage{booktabs}
\usepackage{setspace}
\usepackage{multirow}
\usepackage{epstopdf}
\usepackage{mathtools}
\usepackage{mathdots}
\usepackage{pdfsync}

\newtheorem{remark}{\textbf{Remark}}
\newtheorem{definition}{\textbf{Definition}}

\newtheorem{lemma}{\textbf{Lemma}}
\newtheorem{example}{\textbf{Example}}
\newtheorem{corollary}{\textbf{Corollary}}
\newtheorem{proposition}{\textbf{Proposition}}

\begin{document}

\title{Design of Quantum Stabilizer Codes From Quadratic Residues Sets}



\author{\IEEEauthorblockN{Yixuan Xie$^1$, Jinhong Yuan$^1$ and Qifu (Tyler) Sun$^2$}\\
\IEEEauthorblockA{${}^1$ The University of New South Wales, Sydney, Australia\\}
\IEEEauthorblockA{${}^2$ University of Science and Technology Beijing, China\\
Email: Yixuan.Xie@student.unsw.edu.au, J.Yuan@unsw.edu.au, qfsun@ustb.edu.cn}
}


\maketitle

\begin{abstract}
We propose two types, namely Type-I and Type-II, quantum stabilizer codes using quadratic residue sets of prime modulus given by the form $p=4n\pm1$. The proposed Type-I stabilizer codes are of cyclic structure and code length $N=p$. They are constructed based on multi-weight circulant matrix generated from idempotent polynomial, which is obtained from a quadratic residue set. The proposed Type-II stabilizer codes are of quasi-cyclic (QC) structure and code length $N=pk$, where $k$ is the size of a quadratic residue set. They are constructed based on structured sparse-graphs codes derived from proto-matrix and circulant permutation matrix. With the proposed methods, we design rich classes of cyclic and quasi-cyclic quantum stabilizer codes with variable code length. We show how the commutative constraint (also referred to as the Symplectic Inner Product constraint) for quantum codes can be satisfied for each proposed construction method. We also analyze both the dimension and distance for Type-I stabilizer codes and the dimension of Type-II stabilizer codes. For the cyclic quantum stabilizer codes, we show that they meet the existing distance bounds in literature.
\end{abstract}

\section{Introduction}

Quantum error correction becomes essential for preserving coherent quantum states against not just unintended quantum transformation, but also other unwanted interactions in quantum communication and computation. With quantum error-correcting codes the original quantum information can be recovered correctly when quantum states of quantum bits, \emph{qubits}, carrying the quantum information are transformed by \emph{quantum noise} \cite{Nielsen:2000book}.
Unlike a classical binary $[N,K]$ code, which protects one of the $2^K$ discrete messages by encoding it into one of the $2^K$ codewords of length $N$, the quantum state of $K$ qubits is specified by $2^K$ complex coefficients. The purpose of designing quantum error-correcting codes is to encode the $K$ qubit state into an $N$ qubit state in such a way that all $2^K$ complex coefficients are perfectly stored and be able to detect and correct errors.

While the importance of quantum error-correcting codes is apparent, initially, many researchers believed the nonexistence of quantum error correction codes and the process of quantum error correction is infeasible. Until the mid 90's, the discovery of the 9-qubit code by Shor \cite{Shor1995} and 7-qubit code by Steane \cite{steane:19962} showed that quantum error correction codes did indeed exist. Their following work in \cite{Calderbank1996,Steane:1996} showed that a class of good quantum codes, namely \emph{Calderbank-Shor-Steane (CSS) codes}, can be constructed from classical linear code $\mathcal{C}$ and its dual-containing code $\mathcal{C}^{\perp}\subset \mathcal{C}$. These findings have led to a rapid evolution in the research of quantum error correction and the design of quantum error-correcting codes. A general theory of quantum error correction and necessary conditions for a quantum system to form a quantum error-correcting code are given in \cite{laf1996,Bennett1996,KnillLaf1997}.

An encoding process of $K$ qubits into $N$ qubits is essentially a linear mapping of $\mathcal{H}^{\otimes K}$ onto a $K$-dimensional subspace of $\mathcal{H}^{\otimes N}$, where $\mathcal{H}^{\otimes N}$ denotes the quantum state space of $N$ qubit under tensor product `$\otimes$'.
A special finite group of \emph{unitary} transformations of $\mathcal{H}^{\otimes N}$, known as a \emph{Clifford group}, contains all the transformations necessary for encoding quantum codes.
Investigation of the connection between \emph{Clifford group} and existing quantum codes has led to a general construction of many new quantum codes. The initial results on this investigation were given in \cite{CalderRainShorSloane1997} and later on in \cite{CalderRainShorSloane1998}. Several standard techniques to construct quantum codes over finite field $\mathbb{F}_2$ and $\mathbb{F}_4$ based on the theory of classical coding are given. Thereafter,
various types of quantum error correcting codes have been proposed including the formalism of  \emph{stabilizer codes} \cite{Gottesman1996}, \cite{GottesmanThesis}, which yielded many useful insights and permitted a great number of new codes discovered using classical codes, for instance, quantum BCH codes \cite{MGrasslBeth1999}\cite{SAly:2007}, quantum Reed-Solomon codes \cite{MGrasslBethRS}, quantum convolutional codes \cite{Almeida2004} \cite{Grassl:2007} \cite{Forney2005}, codeword stabilized quantum codes \cite{Cross2009}, non-binary quantum stabilizer codes \cite{AshKnill2000} and quantum non-additive codes \cite{Rains:1997} \cite{Smolin:2007}.

The conventional \emph{cyclic} codes are good candidates of error-correcting codes in terms of high minimum distance and low encoding complexity \cite{MacWilliams:book1978}. These advantages enable the design of quantum cyclic codes and quantum \emph{shift registers} that have been initially studied in \cite{GrasslBeth2000}. It is known that a conventional $[N, K]$ cyclic code can be fully generated from a unique monic polynomial $G(x)$ of minimal degree $N-K$ over a field $\mathbb{F}_q$. This monic polynomial $G(x)$ is often called the \emph{generator polynomial}. However, the generator polynomial of a cyclic code is not easy to obtain especially in the field of higher order. Moreover, while majority classical cyclic codes attain promising distance property, they tend to have poor sparseness when code length is large.

Fortunately, the conventional sparse-graph codes, particularly low-density parity-check (LDPC) \cite{Gallager1963} codes ascertain both the sparseness and a large minimum distance. A well known subclass of LDPC codes, namely \emph{quasi-cyclic} LDPC codes, also possess the simplicity of encoding. However, compared with the design of LDPC codes in classical settings, the commutative constraint (also referred to as \emph{the symplectic inner product} constraint) for quantum codes that sets on a pair of parity-check matrices complicates the design of quantum LDPC codes. In particular, the classical design of LDPC codes utilizing randomness \cite{macKay1999}-\cite{TenBrinkTOC04} is not helpful in the design of quantum LDPC codes. The idea of quantum LDPC codes was first given by Postol in \cite{PostolLDPC}, whereas generalization of quantum LDPC codes was proposed a few years later by MacKay \emph{et al.} \cite{macKay2004}.
Since then, a wide range of different types of sparse-graph quantum codes have been designed, \emph{e.g.,} \cite{aly2007}-\cite{kenta2011}.

The purpose of the present paper is to investigate efficient methods for designing $[[N,K,d_{min}]]$ quantum stabilizer codes, where the double solid bracket is commonly used to distinguish from a classical $[N,K]$ code. In particular, we focus on the use of \emph{quadratic residue (QR) sets} as our basic building block to explore systematic design techniques for new quantum stabilizer codes. More importantly, by taking full advantages of both conventional cyclic codes and sparse-graph codes, we aim at constructing new structured quantum stabilizer codes of variable code length. To illustrate the potential of our proposed methods, we focus on two particular categories of quadratic residue sets and show that they exhibit different properties under the same commutative constraint.

We propose two types of quantum stabilizer codes of length $N=p$ and $N=pk$, respectively, where $p=4n\pm 1$ is the prime modulus of a quadratic residue set for some non-negative integer $n$ and $k$ is the size of the quadratic residue set. We refer to them as Type-I and Type-II quantum stabilizer codes.
To illuminate the simplicity of the proposed construction methods, unlike traditional cyclic codes that are generated from a generator polynomial,
we design Type-I quantum stabilizer codes from \emph{idempotent polynomial} of quadratic residue sets.
Furthermore, we give systematic design methods for Type-II quasi-cyclic stabilizer (QCS) codes by introducing a two step \emph{position-and-lift} operation. In this design, we first place each quadratic residue set into a special format of \emph{Latin square}, as we refer to as a \emph{proto-matrix}. Then, by using \emph{circulant permutation matrices}, we lift the pre-obtained proto-matrix and obtain the parity-check matrix for a Type-II QCS code.

The key results of the paper, the design methods for Type-I and Type-II quantum stabilizer codes, can be applied to any quadratic residue set of prime modulus $p = 4n\pm 1$. In addition, we prove that the minimum distance for Type-I stabilizer codes of length $N=4n+1$ is upper bounded by the size of quadratic reside set $k$. Moreover, the code rate for Type-I stabilizer codes of length $N=4n-1$ is determined by $\frac{K}{N}=\frac{k}{p}$, and the code rate approaches $\frac{1}{2}$ as $n$ goes to infinity.
Furthermore, in the design of Type-II QCS codes, we use the generator element of quadratic residue sets to construct the proto-matrix. To satisfy the commutative constraint, two proto-matrices need to be \emph{commutative} and \emph{adjunction} of an additional element to the proto-matrix is required for codes with $p=4n-1$, whereas for codes with $p=4n+1$ only one proto-matrix needs to be commutative. By using decomposition of Vandermonde matrix, we show that the dimension of Type-II QCS codes is $k-1$ if $n$ is odd or $2k-1$ if $n$ is even.

The organization of the paper is as follow: we review the theory of quantum stabilizer codes in the next Section. In Section III, we give explicit design procedures and analysis for Type-I stabilizer codes of length $N=4n\pm1$ after the preliminaries on quadratic residue sets and its idempotents. Thereafter, we design Type-II QCS codes from quadratic residue sets of modulus $p=4n\pm 1$. Before we conclude in Section VI with further discussions, we present some constructed codes of both Type-I and Type-II stabilizer codes in Section V.

\section{Quantum stabilizer group and stabilizer codes}
The state of single qubit $\vert\psi\rangle\in\mathcal{H}^{\otimes 1}$ exists in the \emph{superposition} $\vert\psi\rangle=\alpha\vert 0\rangle+\beta\vert 1\rangle$ of two basis states $\vert0\rangle$ and $\vert 1\rangle$, where $\alpha$ and $\beta$ are complex numbers and $\vert\alpha\vert^2+\vert\beta\vert^2=1$. Likewise, the state of $N$ qubit could exist in the form $\alpha_{0}\vert00\ldots0\rangle+\alpha_{1}\vert00\ldots1\rangle+\ldots+\alpha_{2^N - 1}\vert11\ldots1\rangle$, where $\sum_{i=0}^{2^N-1}\vert\alpha_i\vert^2=1$ and $\vert11\ldots1\rangle$ is shorthand for the tensor product $\vert1\rangle\otimes\vert1\rangle\otimes\ldots\otimes\vert1\rangle$. Thus, the quantum state space of $N$ qubits $\mathcal{H}^{\otimes N}$ is a composition of $N$ copies of $\mathcal{H}^{\otimes 1}$ through tensor product, where each copy is corresponds to a qubit. Since the process of \emph{quantum measurement} destroys the superposition state \cite{Nielsen:2000book}, quantum error-correcting codes of stabilizer framework ensure that with partial measurement outcome from \emph{unitary operators}, reliable transmission of encoded states can be achieved. In the rest of the section, a brief overview of quantum stabilizer codes is given.
\subsection{Stabilizer group and stabilizer codes}
Let $\mathcal{H}^{\otimes N}=\{\vert\psi\rangle\}$ be the quantum state space of $N$ qubits and $\mathcal{P}_1=i^c\{I=[\begin{smallmatrix}1&0\\0&1\end{smallmatrix}],X=[\begin{smallmatrix}0&1\\1&0\end{smallmatrix}],Z=[\begin{smallmatrix}1&0\\0&-1\end{smallmatrix}],Y=iXZ\}$ be the Pauli group that acts on a single qubit, where $i=\sqrt{-1}$ is the imaginary unit and $i^c$ is the overall phase factor with $c\in\{0, 1, 2, 3\}$. {\color{black} The non-identity operators that act on a single qubit imply} \emph{bit error} (X), \emph{phase error} (Z) or both (Y), respectively. Then the $N$-fold tensor product $\left(\otimes\right)$ of $\mathcal{P}_1$ forms an $N$-qubit Pauli group denoted as
\begin{align}
\mathcal{P}_N = i^c\left\{I,X,Y,Z\right\}^{\otimes N}.
\end{align}
An element in $\mathcal{P}_{N}$ that acts on a state $\vert\psi\rangle\in\mathcal{H}^{\otimes N}$ can be expressed as an operator $E = i^{c}A^{1}\otimes A^{2}\otimes\ldots\otimes A^{N}$, where each $A^{j}\in\mathcal{P}_1$ for $1\leq j\leq N$.  For simple representation, $E = i^{c}A^{1} A^{2}\ldots A^{N}$ and only the non-identity operators $\{X,Z,Y\}$ are identified, \emph{e.g.,} $IXXZZ$ is denoted as $X^{2}X^{3}Z^{4}Z^{5}$.

Any two operators of $\mathcal{P}_N$ are either \emph{commute} or \emph{anti-commute}. For two Pauli operators $E,F\in\mathcal{P}_{N}$, we have
\begin{align}
\label{equ:commutivity}
E\circ F &= +1 \hspace{0.5cm} \text{if $EF = F E$}, \notag\\
E\circ F &= -1 \hspace{0.5cm} \text{if $EF = -F E$},
\end{align}
where $`\circ '$ represents the commutativity between two operators.
Two operators commute if their product shows an $+1$ eigenvalue, otherwise, they  anti-commute.
Furthermore, every element $E\in\mathcal{P}_N$ squares to $\pm 1$ eigenvalue.

A stabilizer group $\mathcal{S}$ is an \emph{Abelian} subgroup of $\mathcal{P}_N$ such that a non-trivial subspace $\mathcal{C_S}$ of $\mathcal{H}^{\otimes N}$ is fixed (or stabilized) by $\mathcal{S}$. The subspace $\mathcal{C}_{\mathcal{S}}$ defines a quantum code space such that
\begin{equation*}
\mathcal{C}_{\mathcal{S}} = \{|\psi\rangle\in\mathcal{H}^{\otimes N} |\hspace{0.3cm} g|\psi\rangle = |\psi\rangle, \forall g\in\mathcal{S}\}.
\end{equation*}
If $\mathcal{S}$ is generated by $\mathbf{g}= \{g_1, g_2,\ldots, g_m\}$, where $\bf{g}$ is the $m = N-K$ independent stabilizer generators, the code space $\mathcal{C}_{\mathcal{S}}$ encodes $K$ logical qubits into $N$ physical qubits and it is able to correct $t=\lfloor\frac{d_{min}-1}{2}\rfloor$ errors. This code $\mathcal{C_S}$ is called an $[[N, K, d_{min}]]$ quantum stabilizer code. Note that a quantum stabilizer code $\mathcal{C_S}$ has the following features:
\begin{align}
1)\hspace{0.2cm}&-I\notin\mathcal{S}, \notag\\
2)\hspace{0.2cm}&\text{for any two} \hspace{0.2cm}E,F\in\mathcal{S}, \hspace{0.2cm} E\circ F=+1.
\end{align}
These properties show that $\mathcal{S}$ needs to be a $+1$ eigenspace spanned by $\mathbf{g}$ (real operators of $\mathcal{P}_N$ are considered only), and $\mathcal{S}$ contains only commuting operators.

Let $\mathbb{E}\subset\mathcal{P}_N$ be a collection of Pauli operators. The condition for quantum error correction in \cite{CalderRainShorSloane1997} \cite{GottesmanThesis} is that $\mathbb{E}$ is a set of correctable error operators for $\mathcal{C_S}$ if
\begin{align}
E^{\dag} F\notin\mathcal{N}(\mathcal{S})\backslash\mathcal{S}, \hspace{0.2cm} \forall E,F\in\mathbb{E},
\end{align}
where `$\dag$' is the conjugate transpose of $E$ and $\mathcal{N}(\mathcal{S})$ the \emph{normaliser} of $\mathcal{S}$ in $\mathcal{P}_N$ such that
\begin{equation*}
\label{secIIB}
\mathcal{N}(\mathcal{S}) = \{A\in\mathcal{P}_N | A^{\dag} EA \in \mathcal{S}, \forall E\in\mathcal{S}\}.
\end{equation*}
Note that $\mathcal{N}(\mathcal{S})$ is a collection of all operators in $\mathcal{P}_N$ that commutes with $\mathcal{S}$ and $\mathcal{S}\subset\mathcal{N}(\mathcal{S})$. Then the minimum distance $d_{min}$ of a stabilzier code is given by
\begin{align}
d_{min}=min(wt(E)) \hspace{0.2cm}\mathbf{s.t.}\hspace{0.2cm} E\in\mathcal{N}(\mathcal{S})\backslash\mathcal{S},
\end{align}
where the \emph{weight}, $wt(*)$, of an operator is the number of positions not equal to Pauli operator $I$. If the stabilizer group $\mathcal{S}$ contains element of weight less than $d_{min}$, then it is a \emph{degenerate} quantum stabilizer code, otherwise, it is a \emph{non-degenerate} quantum stabilizer code.

To correct errors of weight $t$ or less, the particular error operator is determined by measuring the set of stabilizer generators $\mathbf{g}$. The measurement outcome, the \emph{error syndrome}, is a list of eigenvalues of length $m$ denote as $\mathcal{M}(E)=\{+1,-1\}^{m}$.
When an error $E\in \mathcal{P}_N$ acts on the state $\left|\psi\right\rangle\in\mathcal{C_S}$, the corrupted state $E\left|\psi\right\rangle$ is either in $\mathcal{S}$ or anti-commutes with some operators in the $\mathcal{S}$. This is because any two operators of $\mathcal{P}_N$ commute or anti-commute from (\ref{equ:commutivity}). For a non-degenerate stabilizer code, the error syndrome is unique for every correctable errors $E$, whereas for a degenerate stabilizer code, the error syndrome is not unique. Let $E,F\in\mathcal{P}_N$, the former type of stabilizer codes distinguishes $E$ from $F$ if $EF$ anti-commutes with some elements of $\mathcal{S}$, and the later type cannot since $EF\in\mathcal{S}$, \emph{i.e.,} $E$ and $F$ act in the same way on the state.

\subsection{Binary domain and check matrices}
It is known that any stabilizer code can be represented in binary domain $\mathbb{F}_2$. Define the mapping $\Phi : \mathcal{P}_N\rightarrow\mathbb{F}_2^{2N}$. Then $\Phi\left(\mathcal{P}_1\right)=\{(0|0), (1|0), (0|1), (1|1)\}$. This implies that
any operator $E\in\mathcal{P}_N$ can be uniquely expressed as a binary $2N$-tuples obtained through the mapping, that is
\begin{align*}
\Phi(E) = i^{\bar{c}}(\mathbf{a} |{\mathbf{b}}),
\end{align*}
where $\mathbf{a}, \mathbf{b}\in \mathbb{F}_2^{N}$ are two binary $N$-tuples and $i^{\bar{c}}$ is the overall phase factor with $\bar{c}\in\{0, 1, 2, 3\}$. In this representation, $a_j=1$ indicates a bit-error on qubit $j$, $b_j=1$ indicates a phase error on qubit $j$, both errors on the same qubit is represented by $a_j=b_j=1$. For example,
\begin{align}
\label{equ:exampleOperators}
\begin{array}{*{20}{l}}
E &= XYYZI \rightarrow \Phi(E)={(1 1 1 0 0|0 1 1 1 0)},\\
F &= XYZYY \rightarrow \Phi(F)={(11011|01111)}.
\end{array}
\end{align}
Since for any stabilizer code $\mathcal{C_S}$, $-I\notin\mathcal{S}$, we ignore the overall phase factor $i^{\bar{c}}$ in our consideration.

For any two operators $E=(\mathbf{a}|{\mathbf{b}})$ and $F=(\mathbf{a'}|{\mathbf{b'}})$, where $(\mathbf{a}|\mathbf{b})$ and $(\mathbf{a'}|\mathbf{b'})$ are two distinct binary $2N$-tuples, we know that they must commute or anti-commute, that is $EF=\pm FE$. The sign is determined by $(-1)^{(\mathbf{a}\cdot \mathbf{b'})+(\mathbf{a'}\cdot \mathbf{b})}$, where $(\mathbf{a\cdot b')+(a'\cdot b})$ is known as the commutative constraint or twisted inner product. Hence, two elements commute \emph{iff} their corresponding $2N$-tuple $\mathbf{(a|b)}$ and $\mathbf{(a'|b')}$ satisfies the commutative constraint
\begin{align}
\label{equ:twistinner}
 \mathbf{(a\cdot b')+(a'\cdot b)} \equiv 0\left( \bmod 2\right),
\end{align}
where `$\cdot$' is the usual dot product $(\mathbf{a\cdot b'})=\sum_j a_j b'_j$. Using the same example from (\ref{equ:exampleOperators}), the two operators $E$ and $F$ are commuting pairs since $(11100)\cdot(01111)+(11011)\cdot(01110) \equiv 0\left(\bmod 2\right)$. Denoted by $wt(\mathbf{a|b})$, the weight of an operator $(\mathbf{a|b})=(a_1,a_2,\ldots,a_n|b_1,b_2,\ldots,b_n)\in\mathbb{F}_{2}^{2N}$ is the number of positions $j$ such that at least one of $a_j$ and $b_j$ is $1$.

For a stabilizer group $\mathcal{S}$ generated from $m$ independent stabilizer generators $\mathbf{g} = \left\{g_1, g_2,\ldots,g_m\right\}$, define the parity-check matrix $H$ of $\mathcal{S}$ by representing each row of $H$ as $\Phi(g_j)$ for $1\leq j\leq m$ and $g_j\in\mathbf{g}$, the resulting $H$ of size $m\times 2N$ is of the form $H=\left[H_1|H_2\right]$, where
\begin{align}
H_1=\left[\begin{matrix}{\mathbf{a}_{g_1}}\\{\mathbf{a}_{g_2}}\\\vdots\\{\mathbf{a}_{g_m}}\end{matrix}\right] \hspace{0.2cm}\text{and}\hspace{0.2cm} H_2=\left[\begin{matrix}\mathbf{b}_{g_1}\\\mathbf{b}_{g_2}\\\vdots\\\mathbf{b}_{g_m}\end{matrix}\right].
\end{align}
Let $h_i=(\mathbf{a}_{g_i}|\mathbf{b}_{g_i})$ and $h_{i'}=(\mathbf{a}_{g_{i'}}|\mathbf{b}_{g_{i'}})$ be two rows of $H$, where $1\leq i,i'\leq m$ and $i\neq i'$. Since any two elements of $\mathcal{S}$ must commute, $h_i$ and $h_{i'}$ must satisfy the commutative condition given in (\ref{equ:twistinner}). This implies that for $m$ independent stabilizer generators to be commutative, the following constraint, called \emph{Symplectic Inner Product} \cite{macKay2004}, must be satisfied:
\begin{equation}
\centering
\label{equ:symplecticconstraint}
H_{1}H_{2}^{T}+H_{2}H_{1}^{T} = \mathbf{0}^{m\times m} \hspace{0.3cm}(\text{mod}\hspace{0.1cm} 2),
\end{equation}
where $\mathbf{0}^{m\times m}$ is a zero matrix and $`T'$ denotes the transpose of a matrix. We call (\ref{equ:symplecticconstraint}) the SIP constraint for quantum stabilizer codes hereafter.

Since $Y=XZ$, denote by $\mathcal{M}_{\mathrm{X}}(E)$ and $\mathcal{M}_{\mathrm{Z}}(E)$ the two binary $m$-tuple error syndromes measured by $H_1$ and $H_2$, respectively, then $\mathcal{M}(E) = \left(\mathcal{M}_{\mathrm{X}}(E)+\mathcal{M}_{\mathrm{Z}}(E)\right)\hspace{0.1cm} (\bmod \hspace{0.1cm}2)$ if we map eigenvalues $+1\rightarrow 0$ and $-1\rightarrow 1$. This indicates that the columns of parity check matrix $H =[ H_1|H_2]$ are error syndromes for error operator $E$ with $wt(E)=1$. Furthermore, consider two different error operators $E,F\in\mathcal{P}_N$ with $wt(E)=wt(F)=1$, a stabilizer code cannot distinguish these two error operators if their product commutes with $\mathcal{S}$. That is, two error operators have the same error syndrome; $\mathcal{M}(E)=\mathcal{M}(F)$ \emph{iff} $\mathcal{M}(EF)=\mathbf{0}^{m}$.

\subsection{Encoding of General Stabilizer Code}
The linear combinations among rows of parity-check matrix $H$ generate the stabilizer group $\mathcal{S}$ in binary modulo-2 addition. Since the dual-space of $H$ is of dimension $2N-m\equiv\left(2(m+K)-m\right)=m+2K$, the normalizer group $\mathcal{N}(\mathcal{S})$ that commutes with $\mathcal{S}$ can be considered as the dual-space of $\mathcal{S}$ generated by an $(m+2K)\times 2N$ binary matrix. The last $2K$ rows are called \emph{logical operators} $\bar{\textbf{X}}$ and $\bar{\textbf{Z}}$ with $|\bar{\textbf{X}}|=|\bar{\textbf{Z}}|=K$. Note that the choices of $\bar{\textbf{X}}$ and $\bar{\textbf{Z}}$ are non-unique as long as they satisfies
 \begin{align}
 \label{logOPproperty}
 &\bar{X}_i \circ\bar{X}_j= +1, \notag \\
 &\bar{Z}_i \circ\bar{Z}_j= +1, \notag \\
 &\bar{X}_i \circ\bar{Z}_j= +1, \hspace{0.5cm}\text{for}\hspace{0.5cm} i\neq j, \notag\\
 &\bar{X}_i \circ\bar{Z}_j= -1, \hspace{0.5cm}\text{for}\hspace{0.5cm} i=j.
\end{align}

The operation of encoding a general stabilizer code can be described as \cite{GottesmanThesis}
\begin{align}
|x_1,x_2,\ldots,x_K\rangle\rightarrow \left(\prod_{1\leq i\leq m}(I+g_i) \right)\bar X_1^{{x_1}}\bar X_2^{{x_2}}\ldots \bar X_K^{{x_K}}|00\ldots 0\rangle,
\end{align}
where $\bar{X_i}$ is the encoded $X$ operator on the $i$-th qubit. The state $|x_1,x_2,\ldots,x_K\rangle$ is a quantum codeword. The binary $K$-tuples $[x_1,x_2,\ldots,x_K]$ represent one of the $2^K$ possible basis states that can be encoded into. Since a $Z$ operator does not generally affects the basis of a state, only $\bar{X}_i$ operators are used during the encoding process. Recall that the choice of $\bar{\textbf{X}}$ is non-unique. One way to obtain a set of $K$ logical operators $\bar{\textbf{X}}$ or $\bar{\textbf{Z}}$ is to transform the parity-check matrix $H$ of $\mathcal{S}$ into \emph{standard form} \cite{GottesmanThesis}. Hence,
\begin{align*}
\label{logicOp}
{H_{std}} =
\begin{array}{r}
{{R_{{H_1}}} \{ } \\
{m - {R_{{H_1}}}\{ }
\end{array}
\left( {\begin{array}{c|c}
{\underbrace {\begin{array}{*{20}{c}}
I\\
0
\end{array}}_{{R_{{H_1}}}}\underbrace {\begin{array}{*{20}{c}}
{{A_1}}\\
0
\end{array}}_{m - {R_{{H_1}}}}\underbrace {\begin{array}{*{20}{c}}
{{A_2}}\\
0
\end{array}}_K}&{\underbrace {\begin{array}{*{20}{c}}
B\\
D
\end{array}}_{{R_{{H_1}}}}\underbrace {\begin{array}{*{20}{c}}
{{C_1}}\\
I
\end{array}}_{m - {R_{{H_1}}}}\underbrace {\begin{array}{*{20}{c}}
{{C_2}}\\
E
\end{array}}_K}
\end{array}} \right),
\end{align*}
where $R_{{H_1}}$ is the rank of $H_1$. To satisfy conditions in (\ref{logOPproperty}), we obtain $\bar{\textbf{X}}$ and $\bar{\textbf{Z}}$ as
\begin{align}
\bar{\bf{X}} = K\{ ( {\underbrace 0_{{R_{{H_1}}}}\underbrace {{E^T}}_{m - {R_{{H_1}}}}\underbrace I_K|\underbrace {C_2^T}_{{R_{{H_1}}}}\underbrace 0_{m - {R_{{H_1}}}}\underbrace 0_K} )
\end{align}
and
\begin{align}
\bar{\bf{Z}}=K\{ ( {\underbrace 0_{{R_{{H_1}}}}\underbrace 0_{m - {R_{{H_1}}}}\underbrace 0_K|\underbrace {A_2^T}_{{R_{{H_1}}}}\underbrace 0_{m - {R_{{H_1}}}}\underbrace I_K} ),
\end{align}
respectively.

\section{Design of Type-I Quantum Stabilizer Codes}
In this section, we design Type-I quantum stabilizer codes over the finite field $\mathbb{F}$ of order two by exploiting the notion of quadratic reside sets.
The rest of the section is organized in the following way. A preliminary on quadratic residue sets and its idempotents is first introduced. We then design Type-I stabilizer codes for code length $N=4n\pm1$.
Hereafter, we denote the rank of a matrix as $Rank(*)$, the dimension of a code as $dim(*)$ and the degree of a polynomial as $deg(*)$.

\subsection{Quadratic (Non-) Residue Sets and Idempotent Polynomials}

Let $\mathcal{G}_{\mathbb{Z}_p}^{\times}$ be a multiplicative group of order $p$, where $p$ is a prime of the form $p=4n\pm 1$. Denoted by $\mathcal{Q^R}$ and $\mathcal{Q^{NR}}$ the quadratic residue set and quadratic non-residue set, respectively. Take $\alpha$ as a primitive element in $\mathbb{F}_{p}$. Then we have the following.
\begin{lemma}
\label{lem:QRset}
$\mathcal{Q^R} =\{\alpha^{2i} |1\leq i\leq \frac{p-1}{2}\}$ and $\mathcal{Q^{NR}} =\{\alpha^{2i-1} |1\leq i\leq \frac{p-1}{2}\}$ with $\vert\mathcal{Q^R}\vert = \vert\mathcal{Q^{NR}}\vert=\frac{p-1}{2}$.
$\blacksquare$
\end{lemma}

From \emph{Lemma \ref{lem:QRset}}, we know that $\mathcal{Q^R}\bigcup\mathcal{Q^{NR}}=\mathcal{G}_{\mathbb{Z}_p}^{\times}$ since there are exactly half odd and half even integer numbers in $\mathcal{G}_{\mathbb{Z}_p}^{\times}$. Furthermore, for $1\leq i,i'\leq\frac{p-1}{2}$ and $i\neq i'$,  $\alpha^{2i}\cdot\alpha^{2i'}\equiv\alpha^{2i-1}\cdot\alpha^{2i'-1}=\alpha^{0(\bmod 2)}\in\mathcal{Q^R}$ and $\alpha^{2i-1}\cdot\alpha^{2i'}=\alpha^{1(\bmod 2)}\in\mathcal{Q^{NR}}$. We have the following property as a direct consequence of \emph{Lemma \ref{lem:QRset}}.
\begin{lemma}
For $1\leq i\leq\frac{p-1}{2}$,
\label{lem:QRSproperty}
\begin{align*}
&\alpha^{2i}\mathcal{Q^R}=\alpha^{2i-1}\mathcal{Q^{NR}}\equiv\mathcal{Q^R}, \notag \\
&\alpha^{2i-1}\mathcal{Q^R}=\alpha^{2i}\mathcal{Q^{NR}}\equiv\mathcal{Q^{NR}}. \blacksquare
\end{align*}
\end{lemma}

Let $\mathcal{{\bar Q}^{R}} = \{0, \mathcal{Q^{NR}}\}$ and $\mathcal{{\bar Q}^{NR}} = \{0, \mathcal{Q^{R}}\}$ be the \empty{complementary set} of $\mathcal{Q^R}$ and $\mathcal{Q^{NR}}$, respectively. Then for each $\mathcal{G}_{\mathbb{Z}_p}^{\times}$, we can construct four cyclic codes $\mathcal{C}_R$, $\mathcal{\bar{C}}_R$, $\mathcal{C}_{NR}$ and $\mathcal{\bar{C}}_{NR}$ associated to $\mathcal{Q^R}$, $\mathcal{\bar{Q}^R}$, $\mathcal{Q^{NR}}$ and $\mathcal{\bar{Q}^{NR}}$, respectively.
One way to obtain a generator matrix for these codes is to use their \emph{idempotent polynomial}.
Define $\{\mathbb{Q}^r(x), \mathbb{\bar{Q}}^r(x), \mathbb{Q}^{nr}(x), \mathbb{\bar{Q}}^{nr}(x)\}\in\mathbb{F}_2[x]/(x^p-1)$
the idempotent polynomial for $\mathcal{C}_R$, $\mathcal{\bar{C}}_R$, $\mathcal{C}_{NR}$ and $\mathcal{\bar{C}}_{NR}$ over $\mathbb{F}_2$ of a prime $p$. Then
\begin{align}
\label{idempotent}
&\mathbb{Q}^r(x)=\sum_{i\in\mathcal{Q^R}} x^i, \hspace{0.5cm} &\mathbb{\bar{Q}}^r(x)=1 + \sum_{i\in\mathcal{Q^{NR}}} x^{i}, \notag \\
&\mathbb{Q}^{nr}(x)=\sum_{i\in\mathcal{Q^{NR}}} x^{i}, \hspace{0.5cm} &\mathbb{\bar{Q}}^{nr}(x)=1 + \sum_{i\in\mathcal{Q^{R}}} x^{i}.
\end{align}

Let $P$ be the $p\times p$ \emph{circulant permutation matrix} (CPM)
\begin{align}
\label{equ:cpm}
\setlength{\arraycolsep}{4.0pt}
\renewcommand{\arraystretch}{0.4}
P = \left[ {\begin{matrix}
0&1&0& \cdots &0\\
0&0& 1 & \ddots & \vdots \\
 \vdots & \ddots & \ddots & \ddots &0\\
0& \ddots & \ddots & \ddots &1\\
1&0& \cdots & 0 & 0
\end{matrix}} \right].
\end{align}
The generator matrix for $\mathcal{C}_R$ is obtained as
\begin{align}
\mathbb{Q}^r(P) = \sum_{i\in\mathcal{Q^R}} P^i,
 \end{align}
where the $i$-th power of $P$ is the $i$-th cyclic shift of $P$, and $P^0=I$ is the identity matrix.  The transpose of $\mathbb{Q}^r(x)$ is then given by $\mathbb{Q}^r(x^{-1})$. Hence, in matrix representation, it is equivalent to
\begin{align}
\mathbb{Q}^r(P)^T = \sum_{i\in\mathcal{Q^R}} P^{-i}.
 \end{align}
Since $\mathcal{C}_R$ is a cyclic code, where each row of $\mathbb{Q}^r(P)$ is a cyclic shift of previous row by one position, $\mathbb{Q}^r(P)$ can be completely characterized in its idempotent polynomial. Similar representations are used for $\mathcal{\bar{C}}_R$, $\mathcal{C}_{NR}$ and $\mathcal{\bar{C}}_{NR}$.

\subsection{Type-I Stabilizer codes of length $N=4n-1$}
We now look at Type-I stabilizer codes of length $N=4n-1$ by designing multi-weight circulant matrices $H_1$ and $H_2$ from idempotent polynomials in (\ref{idempotent}) first. Then, we analyse the dimension of Type-I stabilizer codes by constructing a pair of sub-matrices $H_1^{sub}$ and $H_2^{sub}$ from $H_1$ and $H_2$. Moreover, we prove the distance of Type-I codes of length $N=4n+1$ is upper bounded by the size of quadratic residue sets.

\begin{proposition}
\label{prop:typeIb}
For an even $n$ and a prime $p=4n-1$, let $H_{1}(x) = \mathbb{\bar{Q}}^{r}(x)$ and $H_{2}(x) = \mathbb{Q}^r(x)$. Then, there exists a pair of sub-matrices $H_1^{sub}$ and $H_2^{sub}$ such that the parity-check matrix $H = [H_{1}^{sub}|H_{2}^{sub}]$ satisfies the SIP constraint with $Rank(H_{1}^{sub})=p-k-1$ and $Rank(H_{2}^{sub}) = p-k$. The resulting parity-check matrix $H$ is a $[[N,K,d_{min}]]=[[p,k,d_{min}=2]]$ Type-I stabilizer code.
$\blacksquare$
\end{proposition}

\begin{IEEEproof}
When $n$ is even, $p=4n-1\equiv -1 \hspace{0.1cm}(\bmod \hspace{0.1cm}8)$, by the $2^{nd}$ Supplement to the Law of Quadratic Reciprocity \cite{Bach:1996}, $2\in\mathcal{Q^{R}}$ and $-1\notin\mathcal{Q^R}$.
From (\ref{idempotent}), $H_1(x)=\mathbb{\bar{Q}}^{r}(x)=1 + \sum_{i\in\mathcal{Q^{NR}}}x^{i}$ and $H_2(x)=\mathbb{{Q}}^{r}(x)=\sum_{j\in\mathcal{Q^R}}x^{j}$. Since $-\mathcal{Q^R}=\mathcal{Q^{NR}}$ and for $f(x)=(x^a + x^b)\in\mathbb{F}_2[x]$, $f(x)^2=(x^{a}+x^{b})^2=x^{2a}+x^{2b}$, we have
\begin{align}
\label{equ:proofpro2}
H_1(x)H_2(x^{-1})&=\left(\sum_{j\in\mathcal{Q^R}} x^{-j}\right) + \left(\sum_{j\in\mathcal{Q^R}} x^{-j}\right) \left(\sum_{i\in\mathcal{Q^{NR}}} x^{i}\right) \notag \\
&\equiv \left(\sum_{i\in\mathcal{Q^{NR}}} x^{i}\right) + \left(\sum_{i\in\mathcal{Q^{NR}}} x^{2i}\right).
\end{align}
By (\ref{lem:QRSproperty}), $\left(\sum_{i\in\mathcal{Q^{NR}}} x^{2i}\right)=\left(\sum_{i\in\mathcal{Q^{NR}}} x^{i}\right)$ because $2\in\mathcal{Q^R}$. Hence, $H_1(x)H_2(x^{-1})=\mathbf{0}\hspace{0.1cm}(\bmod\hspace{0.1cm}2)$.
Similarly, $H_2(x)H_1(x^{-1})\equiv\mathbf{0}\hspace{0.1cm}(\bmod\hspace{0.1cm}2)$ implies that $H_1$ and $H_2$ are commuting pairs for even $n$.

Since $\mathcal{Q^R}\bigcup\mathcal{Q^{NR}}=\mathcal{G}_{\mathbb{Z}_p}^{\times}$.
Then $H_1$ and $H_2$ are complementary matrices, that is
\begin{align}
\label{equ:allone}
H_1+H_2 = \mathbb{I}^{p\times p},
\end{align}
where $\mathbb{I}^{p\times p}$ is an all-one matrix of size $p\times p$. Since $\mathcal{M}(E) =\left( \mathcal{M}_X(E)+\mathcal{M}_Z(E)\right)\hspace{0.1cm}(\bmod\hspace{0.1cm}2)$, we have $\mathcal{M}(Y^1)=\mathcal{M}(Y^2)=\ldots =\mathcal{M}(Y^{p})=[1,1,\ldots,1]^{m}$. Thus, this code cannot distinguish two single weight $Y$ operators acting on different qubits. Hence, $d_{min}=2$.
\end{IEEEproof}

The rank of $H = [H_{1}^{sub}|H_{2}^{sub}]$ constructed from \emph{Proposition \ref{prop:typeIb}} is determined from the following lemma.
\begin{lemma}\label{lemma:three}
Let $\mathbb{Q}^r\left(x\right):=\sum_{i=1}^{k} x^{d_i}$ and $\mathbb{\bar{Q}}^r(x):=1 + \sum_{i=1}^k x^{-d_i}$,
where $d_{1,2,\ldots,k}\in \mathcal{Q^R}$. Then for $n$ is even and $p=4n-1$ is a prime,
\begin{align}
Rank(\mathbb{Q}^r (x))&=p-k.\\
Rank(\mathbb{\bar{Q}}^r (x))&=p-(k+1).
\end{align}
$\blacksquare$
\end{lemma}
\begin{IEEEproof}
For simplicity, write $f\left(x\right)=\mathbb{Q}^r\left(x\right)$. Let $\alpha$ be a primitive $p$-th root of unity in some field $\mathbb{F}_{p}$. To prove the lemma, it is equivalent to find the number of roots of $f(x)$ in $\{1, \alpha, \alpha^2,\ldots, \alpha^{p-1}\}$.

Since $p=4n-1\equiv -1 (\bmod \hspace{0.1cm}8)$ when $n$ is even, we shall show that corresponding to each $d_i$ in $\mathcal{Q^R}$, either $\alpha^{d_i}$ or $\alpha^{-d_i}$ is a root of $f(x)$. Since $p$ is not congruent to $1$ modulo $4$, by the $1^{st}$ Supplement to the Law of Quadratic Reciprocity \cite{Bach:1996}, $-1$ is not a quadratic residue, and we have $\{d_1,d_2,\ldots,d_k\}\cup -\{d_1,d_2,\ldots,d_k\} = \mathcal{G}_{\mathbb{Z}_{p}}^{\times}$. Consequently, $\bigcup\limits_{1 \le i \le p - 1} {\{ {\alpha^{{d_i}}},{\alpha^{ - {d_i}}}\} }  = \{ \alpha,{\alpha^1}, \cdots ,{\alpha^{p - 1}}\} $. Hence for all $1\leq j\leq p-1$, $f(\alpha^{d_i})+f(\alpha^{-d_i}) = \sum_{j=1}^{k}\alpha^{d_jd_i} + \sum_{j=1}^{k}\alpha^{-d_jd_i} = \sum_{j=1}^{p-1}\alpha^{jd_i} = 1 + \sum_{j=0}^{p-1}\alpha^{jd_i} = 1$, where the last equality holds due to $\alpha^{d_i}\neq 1$ being a root of $x^p-1=\left(x-1\right)\left(x^{p-1}+x^{p_2}+\ldots+x^{1}+1\right)$. Again, by the $2^{nd}$ Supplement to the Law of Quadratic Reciprocity, $2\in \mathcal{Q^R}$ in $\mathcal{G}_{\mathbb{Z}_{p}}^{\times}$. Then the quadratic residue set $\{d_1, d_2,\ldots, d_k\}$ is closed under multiplication by $2$. As a result, $f(\alpha^{2d_i}) = f(\alpha^{d_i})$. This implies that $f(\alpha^{d_i})$ is an element in $\mathbb{F}_2$. Thus, either $f(\alpha^{d_i})=0$ or $1 - f(\alpha^{-d_i})=f(\alpha^{d_i})=1$. We conclude that either $\alpha^{d_i}$ or $\alpha^{-d_i}$ is a root of $f(x)$. Hence, when $n$ is even, $Rank\left( f(x)\right) = p-k$.

Similarly, if $f(x)=\mathbb{\bar{Q}}^r(x):=1 + \sum_{i=1}^k x^{-d_i}$ for $-d_{1,2,\ldots, k}\in\mathcal{Q^{NR}}$, then $\{0\}\bigcup\{\alpha^{-d_i}|1\leq i\leq k\}$ are the set of roots for $f(x)$. Hence, $Rank(\mathbb{\bar{Q}}^r(x))=p-(k+1)$.
\end{IEEEproof}

\begin{corollary}
\label{coro:T1Nodd}
For an odd $n$ and a prime $p=4n-1$, the parity-check matrix $H = [H_{1}^{sub}|H_{2}^{sub}]$ satisfies the SIP constraint with $Rank(H_{1}^{sub})=p$ and $Rank(H_{2}^{sub}) = p-1$. The resulting parity-check matrix $H$ is a trivial $[[N,K,d_{min}]]=[[p,0,d_{min}]]$ quantum stabilizer code.
$\blacksquare$
\end{corollary}

\begin{IEEEproof}
In this case, $p=4n-1$ is equivalent to $p= 3 \bmod 8$. Denote by $min{(\mathcal{Q^R})}$ the smallest value in $\{d_1,d_2,\ldots,d_k\}$. Then, $f(x)=x^{min(\mathcal{Q^R})}\cdot\sum_{i=1}^{k}x^{(d_i-min(\mathcal{Q^R}))}$. Since $min(\mathcal{Q^R})=1$, there are at most $p-1-min(\mathcal{Q^R})$ non-zero roots of $f(x)$. By the $2^{nd}$ Supplement to the Law of Quadratic Reciprocity, $2\in{\mathcal{Q}^{nr}}$ is a quadratic non-residue in $\mathcal{G}_{\mathbb{Z}_{p}}^{\times}$, hence the order of $2$ in $\mathcal{G}_{\mathbb{Z}_{p}}^{\times}$ is $p-1$. Assume $\alpha^i$ for some $0\leq i\leq p-1$ is also a root of $f(x)$. Since $f(x)$ is a polynomial over field $\mathbb{F}_2$, $f(\alpha^{i\cdot2^j})=f(\alpha^i)^{2^j}=0$ for all $0\leq j\leq p-1$, which implies that there are $p-1$ distinct roots of $f(x)$. But this contradicts to that $f(x)$ has at most $p-1-min(\mathcal{Q^R})< p-1$ non-zero roots. Hence, no roots of $f(x)$ are in the set $\{1, \alpha, \alpha^2,\ldots,\alpha^{p-1}\}$ and $Rank(H_1)=p-K=p$, where $K=0$. By the same argument in \emph{Lemma \ref{lemma:three}}, $Rank(H_{2})=p-1$.
\end{IEEEproof}
Note that the above analysis also applies to the case when $H_1(x)=\mathbb{\bar{Q}}^{nr}(x)$ and  $H_2(x)=\mathbb{Q}^{nr}(x)$.
\begin{example}\label{exp:comp1} For $n=2$, $\mathcal{Q^R} = \{1, 2, 4\}$ and ${\mathcal{\bar{Q}^R}} =  \{0, 3, 5, 6\}$. Let $H_{1}(x) = \mathbb{\bar{Q}}^r(x)$ and $H_{2}(x) = \mathbb{Q}^r(x)$. We have $H(x) = [1 + {x^3} + {x^5} + {x^6}|{x^1} + {x^2} + {x^4}]$ and $Rank(H_1)=p-k-1=3$ and $Rank(H_2)=p-k=4$.
Consider two error operators $E_1,E_2\in\mathcal{P}_N$, where $E_1=IIIYIII$ and $E_2=IIYIIII$, by measuring all four stabilizer generators on each of the operators, we obtain the syndrome $\mathcal{M}(E_1)=[1,1,1,1]^T$ and $\mathcal{M}(E_2)=[1,1,1,1]^T$. Since $\mathcal{M}(E_1)=\mathcal{M}(E_2)$, the code can not distinguish $Y$ errors on arbitrary two qubits. Thus, $2\geq d_{min}$. $\Box$
\end{example}

\subsection{Type-I Stabilizer codes of length $N=4n+1$}
We now look at another Type-I stabilizer codes of structure $H=[H_1|H_2]$ with length $N=4n+1$.
\begin{proposition}
\label{prop:DSSTypeIb}
For an odd $n$ and a prime $p=4n+1$, let $H_{1}(x) = \mathbb{Q}^r(x)$, $H_{2}(x) = \mathbb{Q}^{nr}(x)$, and $H_1^{sub}$, $H_2^{sub}$ be sub-matrices of $H_1$ and $H_2$, respectively. The parity-check matrix $H = [H_{1}^{sub}|H_{2}^{sub}]$ satisfies the SIP constraint with $Rank(H) = Rank(H_{1}^{sub})=Rank(H_{2}^{sub}) = p-1$. The resulting parity-check matrix $H$ yields a $[[N,K,d_{min}]]=[[p,1, d^{\dag} \geq d_{min} \geq 3]]$ quantum stabilizer code, where $d^{\dag}=min(wt(E))$ for $E\in\mathcal{S}$.
$\blacksquare$
\end{proposition}

Since $p=4n+1\equiv 1 \hspace{0.1cm}(\bmod\hspace{0.1cm} p)$, by \emph{Theorem 14} in \cite{Ching2011}, matrices $H_1 = [\mathbb{{Q}}^r(P)]$ and $H_2=[\mathbb{{Q}}^{nr}(P)]$ are commutating pairs and have rank $p-1$.
Moreover, since $k=\frac{p-1}{2}=2n$, both $H_1$ and $H_2$ are even weight circulant matrices. Hence, we have the following lemma.
\begin{lemma}
\label{lem:evencodewords}
For an odd $n>1$ and a prime $p=4n+1$, let $\mathcal{{C}}_{R}$ and $\mathcal{{C}}_{{NR}}$ be two linear cyclic code spanned by $H_1 = [\mathbb{Q}^r(P)]$ and $H_2=[\mathbb{Q}^{nr}(P)]$, respectively. Then $\mathcal{{C}}_{R}$ and $\mathcal{{C}}_{{NR}}$ are linear \emph{even code} that contain codewords of even weight only. For $a\in\mathcal{{C}}_R$ and $b\in\mathcal{{C}}_{NR}$, $min(wt(a))=min(wt(b))=2$.
$\blacksquare$
\end{lemma}
\begin{IEEEproof}
Let $c_1$, $c_2$ be rows of $H_1$, then
\begin{align}
\label{equ:minweight}
wt(c_1+c_2) = wt(c_1)+wt(c_2)-2wt(c_1\cap c_2).
\end{align}
Since $\vert\mathcal{Q^R}\vert=k$, we have $wt(c_1)=wt(c_2)=k$ and $2wt(c_1\cap c_2)=0\hspace{0.1cm}(\bmod \hspace{0.1cm} 2)$. Thus, $wt(c_1+c_2)\equiv 0\hspace{0.1cm}(\bmod \hspace{0.1cm} 2)$. By induction, for any codeword $a\in\mathcal{{C}}_R$, $wt(a)=0\hspace{0.1cm}(\bmod \hspace{0.1cm} 2)$. Let $b$ be any codeword of $\mathcal{{C}}_{NR}$. {\color{black} Similarly, we can also show by induction that $wt(b)=0\left(\bmod \hspace{0.1cm}2\right)$}. Thus, $\mathcal{{C}}_{R}$ and $\mathcal{{C}}_{{NR}}$ are even codes with $wt(a)=wt(b)= 0\hspace{0.1cm} (\bmod \hspace{0.1cm} 2)$.

We know that an even code has a generator polynomial $G(x)$ that is divisible by $(1+x)$. Thus, any $\mathbb{{Q}}^{r}(x)$ over $\mathbb{F}_2$ is divisible by $(1+x)$. Since $Rank(H_1)=Rank(H_2)=p-1$ and $deg((1+x))=1$, $G(x)=1+x$ is the generator polynomial of $\mathcal{{C}}_R$ for any prime length $p=4n+1$ with an odd $n$. Furthermore, the weight of $G(x)$ is $2$. Therefore, for any codeword $a\in\mathcal{{C}}_R$, we have $wt(a)=\{2i|1\leq i \leq \frac{p-1}{2}\}$ and $min(wt(a))=2$.
Similarly, an even code $\mathbb{{Q}}^{nr}(x)$ is also divisible by $G(x)$, which implies that $\mathcal{{C}}_R=\mathcal{{C}}_{NR}$. Hence, the minimum weight of codewords spanned by $H_1$ and $H_2$ is always $2$.
\end{IEEEproof}


By \emph{Lemma \ref{lem:evencodewords}}, we know that $\mathcal{{C}}_R=\langle\mathbb{{Q}}^r(x)\rangle\equiv\langle G(x)\rangle$ and $\mathcal{{C}}_R=\mathcal{{C}}_{NR}$, where $G(x)=1+x$. Let $H_1^{sub}=\left[ G(P)\right]$ and $H_2^{sub}=\left[\mathbb{{Q}}^{nr}(P)\right]$ with rank $p-1$. By linear operation on rows and columns of $H_1^{sub}$ and $H_2^{sub}$, we transform $H=[H_1^{sub}|H_2^{sub}]$ into its \emph{reduced row-echelon} form
\begin{align}
\label{equ:h1equal}
\setlength{\arraycolsep}{1.5pt}
\renewcommand{\arraystretch}{0.6}
{H_{rref}} = \left[ {\begin{array}{c|c}\underbrace{\begin{matrix}
{{I^{\left( {p - 1} \right) \times \left( {p - 1} \right)}}}&{\begin{array}{*{20}{c}}
1\\
1\\
 \vdots \\
1
\end{array}}
\end{matrix}}_{H_1^{sub'}}
&
H_2^{sub'}
\end{array}} \right],
\end{align}
where $H_1^{sub'}$ and $H_2^{sub'}$ are equivalent matrices for $H_1^{sub}$ and $H_2^{sub}$, respectively.
Note that each row of $H_{1}^{sub'}$ is of weight $2$ and the linear combination between any two rows of $H_1^{sub'}$ is also a codeword of weight $2$. Therefore, the total number of weight $2$ codewords ${p \choose 2}$ is the summation of $(p-1)$ and ${p-1\choose 2}$. The corresponding row weight of $H_{2}^{sub'}$ is then determined by the following lemma.
\begin{lemma}
\label{lem:weightH2sub}
Let $c$ be a row of $H_{2}^{sub'}$, where $H_{2}^{sub'}$ is the equivalent matrix of $H_{2}^{sub}$ given in (\ref{equ:h1equal}). Then $min(wt(c))=k$ and $max(wt(c))=k+2$.
$\blacksquare$
\end{lemma}
\begin{IEEEproof}
Let $H_{1}(x) = \mathbb{{Q}}^r(x)$ and $H_{2}(x) = \mathbb{{Q}}^{nr}(x)$. Since $n$ is odd and $p$ is a prime of the form $p=4n+1$, by the $1^{st}$ Supplement to the Law of Reciprocity, $-1\in\mathcal{Q^R}$ and $2\notin\mathcal{Q^R}$. Then by \emph{Lemma \ref{lem:QRSproperty}}, $H_1(x)=H_1{(x^{-1})}$ (resp. $H_2(x)=H_2{(x^{-1})}$) and $H_1(x)^2=H_2(x)$ (resp. $H_2(x)^2 = H_1(x)$), that is, $H_1(x)^2$ (resp. $H_2(x)^2$) it is equivalent to
\begin{align}
\label{equ:h1h1}
H_1H_1^T = kI + (n-1)\mathbb{I}^{p\times p}+H_2\equiv H_2\hspace{0.1cm}(\bmod \hspace{0.1cm}2)
\end{align}
and
\begin{align}
\label{equ:h2h2}
H_2H_2^T = kI + (n-1)\mathbb{I}^{p\times p}+H_1\equiv H_1 \hspace{0.1cm}(\bmod \hspace{0.1cm}2).
\end{align}
It can be seen that the maximum and minimum overlapping between a pair of rows in either $H_{1}$ or $H_{2}$ is $n$ and $n-1$, respectively.
Thus, using (\ref{equ:minweight}), the row weight of $H_{2}^{sub'}$ is equal to $2k-2n=k$ assuming two rows having the maximum overlapping or to $2k-2(n-1)=k+2$ assuming two rows having the minimum overlapping.
\end{IEEEproof}

From above, we have the following result.
\begin{lemma}
\label{coro:weightOP}
Let $E\in\mathcal{S}$ be a Pauli operator of weight $wt(E)$, where $\mathcal{S}$ is the stabilizer group spanned by $H=[H_1^{sub}|H_2^{sub}]$. Let $d^{\dag}=min(wt(E))$ be the minimum weight of operator in $\mathcal{S}$. Then we have
$d^{\dag}\leq k$.
$\blacksquare$
\end{lemma}
\begin{IEEEproof}
From \emph{Lemma \ref{lem:evencodewords}}, we know that $\mathcal{{C}}_R=\mathcal{{C}}_{NR}$. Thus, for any $E\in\mathcal{S}$, $\Phi(E)=(a|b)\in\mathbb{F}_2^{2N}$ with $a,b\in\mathcal{{C}}_R$.
The weight of $E$ is determined by
\begin{align}
\label{equ:opweight}
wt(E)\equiv wt(a|b)=wt(a)+wt(b)-wt(a\cap b).
\end{align}
Then, the minimum weight, $d^{\dag}$, is given by
\begin{align}
\label{equ:minweightddag}
d^{\dag} = min\{wt(a)+wt(b)-wt(a\cap b)\}.
\end{align}
Since $min(wt(a))=2$, Equation (\ref{equ:minweightddag}) is equivalent to
\begin{align}
\label{equ:minweightddag2}
d^{\dag}&=min\left\{ \mathop {min }\limits_{a \in {\mathcal{{C}}_R},wt\left( a \right) = 2}\{wt(a)+wt(b)-wt(a\cap b) \}, \mathop {min }\limits_{a \in {\mathcal{{C}}_R},wt\left( a \right) \neq 2}\{wt(a)+wt(b)-wt(a\cap b) \} \right\} \notag\\
&\leq min(wt(a)) + wt[b|wt(a)=2]  - wt(a\cap b).
\end{align}
We know from \emph{Lemma \ref{lem:weightH2sub}} that $max(wt(b))=k+2$ and $min(wt(b))=k$ when $wt(a)=2$. Therefore,
\begin{align}
\label{equ:minweightddag3}
d^{\dag}
&\leq min(wt(a)) + min[wt(b)|wt(a)=2] - max(wt(a\cap b)) \notag \\
&\leq 2 + k -min\{wt(a), wt(b)\}\equiv k.
\end{align}
\end{IEEEproof}

To encode such a code, note that Equation (\ref{equ:h1equal}) is already in the standard form given in (\ref{logicOp}),
\begin{align}
H_{std}={\left(H_1^{sub'}|\hspace{0.1cm} B\hspace{0.2cm} C\right)},
\end{align}
where $B$ is a $(p-1)\times(p-1)$ square matrix and $C$ is a single $(p-1)\times 1$ column vector. Therefore, the logical operators $\bar{Z}_1$ and $\bar{X}_1$ for $K=1$ are
\begin{align}
\label{equ:x1}
{\bar{Z}_1}=\left(0,0,\ldots,0 |1,1,\ldots,1 \right)
\end{align}
and
\begin{align}
\label{equ:z1}
{\bar{X}_1} = \left(0,0,\ldots,0,1|\hspace{0.1cm} C^T \hspace{0.2cm}0\right).
\end{align}
The minimum distance $d_{min}$ of a stabilizer code that is defined as
\begin{align}
d_{min} = min(wt(E))\hspace{0.2cm} \textbf{s.t.} \hspace{0.2cm}E\in\mathcal{N}(\mathcal{S})\backslash\mathcal{S},
\end{align}
can be determined by the following lemma.

\begin{lemma}
\label{lem:upperboundforDmin}
Let $F\in\mathcal{N}(S)\backslash\mathcal{S}$ be a Pauli operator of weight $wt(F)$. The minimum distance $d_{min}$ is upper bounded by
\begin{align}
d_{min}=min(wt(F))\leq k-1
\end{align}
$\blacksquare$
\end{lemma}
\begin{IEEEproof}
The subset $\mathcal{N}(\mathcal{S})\backslash\mathcal{S}$ is generated by multiplying $\mathcal{S}$ with ${\bar{X}_1}$, $\bar{Z}_1$ and $\bar{X}_1\bar{Z}_1$.
Let $\Phi(\bar{X}_1)=({a}_{\bar{X}_1}|{b}_{\bar{X}_1})$ and $\Phi(\bar{Z}_1)=({a}_{\bar{Z}_1}|{b}_{\bar{Z}_1})$. Let $\Phi(F)=(a'|b')\in\mathbb{F}_2^{2N}$ and $\Phi(E)=(a|b)\in\mathbb{F}_2^{2N}$ be the binary $2N$-tuples for $F\in\mathcal{N}(\mathcal{S})\backslash\mathcal{S}$ and for $E\in\mathcal{S}$, respectively. The binary $N$-tuples $a'$ and $b'$ are determined by one of the linear combinations
\begin{align}
\label{equ:opcombination}
a' &\in \{a+a_{\bar{X}_1}, a+a_{\bar{Z}_1}, a+(a_{\bar{X}_1}+a_{\bar{Z}_1})\}, \notag\\
b' &\in \{b+b_{\bar{X}_1}, b+b_{\bar{Z}_1}, b+(b_{\bar{X}_1}+b_{\bar{Z}_1})\}.
\end{align}
Since $min(wt(b))=k$ given that $min(wt(a))=2$ for $E\in\mathcal{S}$, the weight of the column vector $C$ in (\ref{equ:z1}) is $wt(C)\geq k$. Thus, we have
\begin{align}
\label{equ:logopweight}
wt(a_{\bar{X}_1}) = 1, \hspace{0.4cm}wt(b_{\bar{X}_1}) \geq k,\notag \\
wt(a_{\bar{Z}_1}) = 0, \hspace{0.4cm}wt(b_{\bar{Z}_1}) = p.
\end{align}

The minimum distance $d_{min}$ is given by
\begin{align}
\label{equ:dminupper}
d_{min}&=min(wt(F))\equiv min(wt(a'|b')) \notag \\
&=min\{wt(a')+wt(b')-wt(a'\cap b')\}.
\end{align}
Since either $wt(b)=k+2$ or $wt(b)=k$ given that $min(wt(a))=2$, by considering all the possible cases for the given $wt(b)$ and $min(wt(a))$, Equation (\ref{equ:dminupper}) can be expanded into Equation (\ref{equ:dminbound})
\begin{figure*}[t]
\begin{align}
\label{equ:dminbound}
d_{min}&\leq min\left\{\begin{array}{l}min[wt(a')|wt(a)=2] + wt[b'|min(wt(a')) \hspace{0.1cm}\&\hspace{0.1cm} wt(b)=k] - max(wt(a'\cap b')), \\
min[wt(a')|wt(a)=2] + wt[b'|min(wt(a')) \hspace{0.1cm}\&\hspace{0.1cm} wt(b)=k+2] - max(wt(a'\cap b')), \\
min[wt(b')|wt(b)=k] + wt[a'|min(wt(b')) \hspace{0.1cm}\&\hspace{0.1cm} wt(a)=2] - max(wt(a'\cap b')), \\
min[wt(b')|wt(b)=k+2] + wt[a'|min(wt(b')) \hspace{0.1cm}\&\hspace{0.1cm} wt(a)=2] - max(wt(a'\cap b'))
\end{array}\right\}.
\end{align}
\hrulefill
\end{figure*}

By using (\ref{equ:opcombination}) and (\ref{equ:logopweight}), the upper bound for $d_{min}$ is
\begin{align}
d_{min}&\leq min\left\{k, k+2, k, k-1\right\}\notag\\
&\leq k-1.
\end{align}
We have now completed the proof.
\end{IEEEproof}

The lower bound on the minimum distance $d_{min}$ can be interpreted as the following. Since
\begin{align}
H_{1}(x) + H_2(x) = \mathbb{{Q}}^r(x) + \mathbb{{Q}}^{nr}(x)=\sum_{i=1}^{p-1}x^i,
\end{align}
we have
\begin{align}
H_1+H_2 = \mathbb{{Q}}^r(P) + \mathbb{{Q}}^{nr}(P)=\sum_{i=1}^{p-1}P^i=\mathbb{I}^{\mathcal{G}_{\mathbb{Z}_p}^{\times}}.
\end{align}
Let $E\in\mathcal{N}(\mathcal{S})\backslash\mathcal{S}$ have weight $wt(E)=1$. Then for $1\leq i\leq p$, $\{\mathcal{M}_Z(E_i)|\mathcal{M}_X(E_i)|\mathcal{M}(E_i)\}=[\mathbb{{Q}}^{nr}(P)|\mathbb{{Q}}^r(P)\hspace{0.1cm}$ $\vert \hspace{0.1cm} \mathbb{I}^{\mathcal{G}_{\mathbb{Z}_p}^{\times}}]$ are distinct column vectors of size $p-1$. Hence, $d_{min}\geq 3$ and $H=[H_1^{sub}|H_2^{sub}]$ of length $N=4n+1$ is at least a single error-correctable code.
We now give an example of Type-I stabilizer codes of length $N=4n+1$.
\begin{example}
\label{exp:DSS4np1b}
For $n=3$ and $p=13$, $\mathcal{Q^R}=\{1,3,4,9,10,12\}\left(\bmod \hspace{0.1cm}13\right)$ and $\mathcal{Q^{NR}}=\{2,5,6,7,8,11\}$ $\left(\bmod \hspace{0.1cm}13\right)$. Thus
\begin{align}
\mathbb{{Q}}^r(x) = x+x^3+x^4+x^9+x^{10}+x^{12} \notag
\end{align}
and
\begin{align}
\mathbb{{Q}}^{nr}(x) = x^2+x^5+x^6+x^7+x^{8}+x^{11}.
\end{align}
The rank $Rank(H)=Rank(H_1^{sub})=Rank(H_2^{sub})=13-1=12$. This is a $[[13,1,d_{min}]]$ quantum stabilizer code. The stabilizer and the set of logical operators $\bar{X}_1$ and $\bar{Z}_1$ are shown in TABLE \ref{tab:exp4np1b}. Let $E=g_2g_4\bar{X}_1$. Then $E=IXIYZIIIIIIZY\in\mathcal{N}(\mathcal{S})\backslash\mathcal{S}$ has $wt(E)=5$. Note that $d^{\dag}= 6$ and the minimum distance of this code is $d_{min}=5<d^{\dag}$. Hence, this is a non-degenerate $[[13, 1, 5]]$ stabilizer code that is capable to correct arbitrary two errors.
$\Box$
\end{example}


\begin{table}
\center
\caption{Stabilizer of $[[13,1,5]]$ quantum stabilizer code.}
\label{tab:exp4np1b}
\renewcommand\arraystretch{0.7}
\renewcommand\tabcolsep{4pt}
\begin{tabular}{c|ccccccccccccc}
\toprule
$g_1$ & $X$&$Z$&$Z$&$I$&$Z$&$I$&$I$&$I$&$Z$&$I$&$Z$&$Z$&$X$\\
$g_2$ & $I$&$Y$&$I$&$Z$&$Z$&$Z$&$I$&$I$&$Z$&$Z$&$Z$&$I$&$Y$\\
$g_3$ & $Z$&$Z$&$X$&$I$&$I$&$Z$&$Z$&$I$&$Z$&$Z$&$I$&$I$&$X$\\
$g_4$ & $I$&$I$&$I$&$X$&$Z$&$I$&$Z$&$Z$&$Z$&$Z$&$I$&$Z$&$X$\\
$g_5$ & $I$&$Z$&$Z$&$I$&$Y$&$Z$&$I$&$Z$&$I$&$Z$&$I$&$Z$&$Y$\\
$g_6$ & $Z$&$Z$&$I$&$Z$&$Z$&$Y$&$Z$&$I$&$I$&$I$&$I$&$Z$&$Y$\\
$g_7$ & $Z$&$I$&$I$&$I$&$I$&$Z$&$Y$&$Z$&$Z$&$I$&$Z$&$Z$&$Y$\\
$g_8$ & $Z$&$I$&$Z$&$I$&$Z$&$I$&$Z$&$Y$&$I$&$Z$&$Z$&$I$&$Y$\\
$g_9$ & $Z$&$I$&$Z$&$Z$&$Z$&$Z$&$I$&$Z$&$X$&$I$&$I$&$I$&$X$\\
$g_{10}$ & $I$&$I$&$Z$&$Z$&$I$&$Z$&$Z$&$I$&$I$&$X$&$Z$&$Z$&$X$\\
$g_{11}$ & $I$&$Z$&$Z$&$Z$&$I$&$I$&$Z$&$Z$&$Z$&$I$&$Y$&$I$&$Y$\\
$g_{12}$ & $Z$&$Z$&$I$&$Z$&$I$&$I$&$I$&$Z$&$I$&$Z$&$Z$&$X$&$X$\\
\midrule
$\bar{X}_1$ & $I$&$Z$&$I$&$I$&$Z$&$Z$&$Z$&$Z$&$I$&$I$&$Z$&$I$&$X$\\
$\bar{Z}_1$ & $Z$&$Z$&$Z$&$Z$&$Z$&$Z$&$Z$&$Z$&$Z$&$Z$&$Z$&$Z$&$Z$\\
\bottomrule
\end{tabular}

\end{table}

\section{Type-II Quasi-cyclic Quantum stabilizer codes}

Latin squares are wildly used in the design of Steiner triple systems (STS), which leads to an efficient way to design conventional quasi-cyclic LDPC codes, \emph{e.g.}, \cite{vasic2004} \cite{LiZhang2010}. The general definition of a Latin square is the following.
\begin{definition}
\label{def:latin}
Let $L$ be a set of elements $\{l_1,l_2,\ldots,l_q\}$. An $q\times q$ square matrix
\begin{align}
\label{equ:latin}
S=\left[\begin{matrix}s_{(1,1)} & s_{(1,2)}&\cdots&s_{(1,q)}\\
s_{(2,1)} & s_{(2,2)}&\cdots&s_{(2,q)}\\
\vdots&\vdots&\ddots&\ddots \\
s_{(q,1)} & s_{(q,2)}&\cdots&s_{(q,q)}\\
\end{matrix}\right],
\end{align}
is a Latin square of order $q$ if each row and column of $S$ contains each element of $L$ exactly once. A Latin square is called commutative if cell $(i,j)$ and $(j,i)$ for $1\leq i, j\leq q$ contain the same element of $L$, that is $S = S^T$. $\blacksquare$
\end{definition}

In \cite{aly2007}, a class of quantum LDPC codes of CSS structure has been proposed based on the notion of \emph{Latin Squares}.
We will introduce a type of quantum LDPC codes of quasi-cyclic structure for general stabilizer codes by adopting the two step \emph{position-and-lift} operation.
The first step of the operation is to design a \emph{proto-matrix} by positioning elements of $\mathcal{Q^R}$ (resp. $\mathcal{Q^{NR}}$) into the form given in (\ref{equ:latin}), where each entry of the proto-matrix can be treated as the power of the CPM $P$ given in (\ref{equ:cpm}). By substituting each element at position $(i,j)$ of the proto-matrix with the CPM $P$, we lift the proto-matrix into a square matrix of size $pk\times pk$. The dimension of the associated Type-II quasi-cyclic stabilizer (QCS) codes can be determined thereafter.

In the rest of this section, we first design Type-II QCS codes of length $N=kp$, where $p=4n-1$ and $k=\frac{p-1}{2}$. Then we look at Type-II QCS codes of length $N=kp$ and $p=4n+1$. We name them QCS-A codes and QCS-B codes, respectively.

\subsection{QCS-A codes from QR set of size p=4n-1}

Since $\mathcal{Q^R}$ (resp. $\mathcal{Q^{NR}}$) is a set of elements of size $k$, we first design a $k\times k$ proto-matrix such that each row and column contains each element of $\mathcal{Q^R}$ (resp. $\mathcal{Q^{NR}}$) exactly once. Hence, both proto-matrices are Latin squares of order $k$.
Let $\beta=\alpha^2$, where $\alpha$ is a primitive element of the field $\mathbb{F}_p$. Since $\alpha^2\in\mathcal{Q^R}$ and $k=\frac{p-1}{2}$, $\beta$ is the $k-$th root of unity and $\mathcal{Q^R}$ is closed under multiplication by $\beta$. Thus, $\beta$ is the \emph{generator element} of the $\mathcal{Q^R}$ and we can express $\mathcal{Q^R}$ as
\begin{align}
\mathcal{Q^R} = \{\beta, \beta^{2}, \ldots,\beta^{k}\}.
\end{align}
Moreover, we denote the proto-matrix representation of $\mathcal{Q^R}$ as
\begin{align}
\centering
\label{equ:30}
{h_P}\left( \mathcal{Q^R} \right) = \left[{\mathcal{Q^R}(1)}\hspace{0.1cm}{\mathcal{Q^R}(2)} \hspace{0.1cm}\cdots \hspace{0.1cm}{\mathcal{Q^R}(k)} \right]=\left[{\beta}\hspace{0.1cm}  {\beta^2}\hspace{0.1cm}{\ldots}\hspace{0.1cm}{\beta^k}\right],
\end{align}
where $\mathcal{Q^R}\left(j\right)$, $1\leq j\leq k$, represents the $j$-th element of $\mathcal{Q^R}$. Since $\beta^k=1\left(\bmod\hspace{0.1cm} p\right)$, $\beta^{k+i}\equiv\beta^{i\left(\bmod k\right)}$ for $1\leq i\leq k$. Denote the $i$-th cyclic left shift of ${h_P}\left( \mathcal{Q^R} \right)$ by
\begin{align}
\centering
\label{equ:31}
\begin{array}{l}
circ\left(h_P(\mathcal{Q^R})\right)^i\equiv
{h_P}\left( {{\beta ^i}\mathcal{Q^R}} \right) = \left[
{{\beta ^i}\mathcal{Q^R}( 1 )}\hspace{0.1cm}{{\beta ^i}\mathcal{Q^R}( 2 )} \hspace{0.1cm}\cdots \hspace{0.1cm}{{\beta ^i}\mathcal{Q^R}( k)}\right].\\
\end{array}
\end{align}
Similarly, for set $\mathcal{Q^{NR}}=-\{\beta, \beta^{2}, \ldots,\beta^{k}\}$, we have $\beta^i\mathcal{Q^{NR}}\in\mathcal{Q^{NR}}$, then
\begin{align}
circ\left( h_P(\mathcal{Q^{NR}})\right)^i\equiv
{h_P}\left( {{\beta ^i}\mathcal{Q^{NR}} } \right) = \left[
{{\beta ^i}\mathcal{Q^{NR}}(1)}\hspace{0.1cm} {{\beta ^i}\mathcal{Q^{NR}}(2)}\hspace{0.1cm} \cdots\hspace{0.1cm} {{\beta ^i}\mathcal{Q^{NR}} (k)}\right].
\end{align}
We now construct the proto-matrices ${H_{1}}_{proto}$ and ${H_{2}}_{proto}$ by positioning different shift of $h_P(\mathcal{Q^R})$ and $h_P(\mathcal{Q^{NR}})$ into the following structure
\begin{equation}
\centering
\label{equ:32}
\renewcommand\arraystretch{0.7}
\renewcommand\tabcolsep{4pt}
{H_{1proto}} = \left[ \begin{array}{l}
{h_P}\left( \mathcal{Q^R}\right)\\
circ\left({h_P}\left( {\mathcal{Q^R}}\right) \right)^1\\
\begin{array}{*{20}{c}}
{}& \vdots &{}
\end{array}\\
circ\left({h_P}\left( {\mathcal{Q^R}} \right)\right)^{k-1}
\end{array} \right],{H_{2proto}} = \left[ \begin{array}{l}
{h_P}\left( {\mathcal{Q^{NR}}} \right)\\
circ\left({h_P}\left( { {\mathcal{Q^{NR}}} } \right)\right)^1\\
\begin{array}{*{20}{c}}
{}& \vdots &{}
\end{array}\\
circ\left({h_P}\left( { {\mathcal{Q^{NR}}} } \right)\right)^{k-1}
\end{array} \right].
\end{equation}
Both ${H_{1}}_{proto}$ and ${H_{2}}_{proto}$ are square matrices of $k$ different permutations of $\mathcal{Q^R}$ and $\mathcal{Q^{NR}}$, respectively, where the $i$-th row is the $i$-th permutation. Since the first column of $H_{1proto}$ is $[\beta\hspace{0.2cm} \beta^{1+1}\hspace{0.2cm} \ldots \hspace{0.2cm}\beta^{1+k-1}]^T$, which is equivalent to (\ref{equ:30}), $H_{1proto} = H_{1proto}^T$. Similarly, we have $H_{2proto} = H_{2proto}^T$. Thus, by \emph{Definition \ref{def:latin}}, $H_{1proto}$ and $H_{2proto}$ are commutative Latin squares of order $k$ with every element of $\mathcal{Q^R}$ and $\mathcal{Q^{NR}}$ appearing exactly once in every row and every column. Since $circ\left({h_P}\left( \mathcal{Q^R}\right)\right)^i=-circ\left({h_P}\left( \mathcal{Q^{NR}}\right)\right)^i$ for $0\leq i\leq k-1$,  ${H_{2}}_{proto} = -{H_{1}}_{proto}$ and
$H_{1proto}+H_{2proto} = p\mathbb{I}^{p\times p}$.
Next, we lift each entry of $H_{1proto}$ and $H_{2proto}$ by inserting CPM $P$ of size $p$. For each cyclic shift $0\leq i\leq k-1$ of $h_P(\mathcal{Q^R})$ (resp. $h_P(\mathcal{Q^{NR}})$),  we have
\begin{align}
h_{\mathcal{Q^R}_i} = \left[P^{\beta^{1+i}} \hspace{0.2cm} P^{\beta^{2+i}}\hspace{0.2cm}\ldots \hspace{0.2cm}P^{\beta^{k+i}}\right]
\end{align}
and
\begin{align}
h_{\mathcal{Q^{NR}}_i} = \left[P^{-\beta^{1+i}} \hspace{0.2cm} P^{-\beta^{2+i}}\hspace{0.2cm}\ldots \hspace{0.2cm}P^{-\beta^{k+i}}\right].
\end{align}
Finally, let
\begin{align}
\renewcommand\arraystretch{0.7}
\renewcommand\tabcolsep{4pt}
H_1=\left[\begin{array}{l}h_{\mathcal{Q^R}_0}\\h_{\mathcal{Q^R}_1}\\\vdots\\ h_{\mathcal{Q^R}_{k-1}}\end{array}\right] \hspace{0.5cm}\text{and} \hspace{0.5cm} H_2=\left[\begin{array}{l}h_{\mathcal{Q^{NR}}_0}\\h_{\mathcal{Q^{NR}}_1}\\\vdots\\ h_{\mathcal{Q^{NR}}_{k-1}}\end{array}\right].
\end{align}
We obtain a pair of matrices $H_1$ and $H_2$ of size $pk\times pk$.
Note that each one of $h_{\mathcal{Q^R}_i}$ (resp. $h_{\mathcal{Q^{NR}}_i}$) is a $p\times kp$ sub-matrix of $H_1$ (resp. $H_2$). So we call each  $h_{\mathcal{Q^R}_i}$ (resp. $h_{\mathcal{Q^{NR}}_i}$) a \emph{circulant array}.

Define $\boxplus$ the operation of \emph{adjunction}; \emph{e.g.,} $(3\boxplus 5) \equiv (P^{3} + P^{5})$. Equivalently, it can also be represented as $(x^{3} + x^{5})$ in the polynomial form.
Let $\mathbf{\boldsymbol\emptyset}_{proto}^{1 \times k}$ be an all-zero proto-matrix of size $1\times k$. We adjunct each element of $H_{2proto}$ with element $0$. Thus, the final proto-matrix $H_{proto}$ is of the form
\begin{align}
\label{equ:protoQCDSS}
\renewcommand\arraystretch{0.8}
\renewcommand\tabcolsep{6pt}
\begin{array}{l}
{H_{proto}} = \left[ {{H_{1proto}}|H_{2proto}^{'}} \right] = \\
\left[ {\begin{array}{c|c}{\underbrace {\begin{array}{*{20}{l}}
{{ {{h_P}\left( {\mathcal{Q^R}} \right)}}}\\
{circ{{\left( {{h_P}\left( {\mathcal{Q^R}} \right)} \right)}^1}}\\
{\begin{array}{*{20}{c}}
{}& \vdots &{}
\end{array}}\\
{circ{{\left( {{h_P}\left( {\mathcal{Q^R}} \right)} \right)}^{k - 1}}}
\end{array}}_{H_{1proto}}} & \underbrace{\begin{array}{*{20}{l}}
{{{ {{h_P}\left( {\mathcal{Q^{NR}}} \right)}}}}\\
{circ{{\left( {{h_P}\left( {\mathcal{Q^{NR}}} \right)} \right)}^1}}\\
{\begin{array}{*{20}{c}}
{} & \vdots & {}
\end{array}}\\
{circ{{\left( {{h_P}\left( {\mathcal{Q^{NR}}} \right)} \right)}^{k - 1}}}
\end{array}\begin{array}{*{20}{c}}
{\boxplus\left( {\boldsymbol\emptyset _{proto}^{1 \times k}} \right)}\\
{\boxplus\left( {\boldsymbol\emptyset _{proto}^{1 \times k}} \right)}\\
 \vdots \\
{\boxplus\left( {\boldsymbol\emptyset _{proto}^{1 \times k}} \right)}
\end{array}}_{H_{2proto}^{'} = {H_{2proto}} \boxplus \boldsymbol\emptyset _{proto}^{k \times k}}\end{array}} \right].
\end{array}
\end{align}

\begin{proposition}
\label{prop:2}
For a positive integer $n$ and a prime $p=4n-1$, let $h_P(\mathcal{Q^R})$ and $h_P(\mathcal{Q^{NR}})$ be the proto-matrices of $\mathcal{Q^R}$ and $\mathcal{Q^{NR}}$, respectively. Let $\boldsymbol\emptyset_{proto}^{k \times k}$ be an all-zero proto-matrix of size $k\times k$. The parity-check matrix $H$ generated from the proto-matrix $H_{proto} = \left[ {{H_{1proto}}|{H_{2proto}}} \boxplus \boldsymbol\emptyset_{proto}^{k \times k}\right]$ always satisfies the SIP constraint, that is, ${H_{1}}{H_{2}}^T + {H_{2}}{H_{1}}^T=2\left(\mathbb{I}_{proto}^{k \times k} \bigotimes {\mathbb{I}^{\mathcal{G}_{\mathbb{Z}_p}^{\times}}}\right) \equiv \mathbf{0}^{pk\times pk}\left(\bmod\hspace{0.1cm}2\right)$.
$\blacksquare$
\end{proposition}

\begin{IEEEproof}
Let $H_1$ and $H_2$ be in the form of
\begin{align}
\label{equ:42}
\setlength{\arraycolsep}{1.5pt}
\renewcommand{\arraystretch}{0.6}
{H_1}(x) = \left( {\begin{array}{*{20}{c}}
{{x^\beta }}& \cdots &{{x^{{\beta ^{k - 1}}}}}&{{x^{{\beta ^k}}}}\\
{{x^{{\beta ^2}}}}& \cdots &{{x^{{\beta ^k}}}}&{{x^{{\beta ^1}}}}\\
 \vdots & {\mathinner{\mkern2mu\raise1pt\hbox{.}\mkern2mu
 \raise4pt\hbox{.}\mkern2mu\raise7pt\hbox{.}\mkern1mu}} & {\mathinner{\mkern2mu\raise1pt\hbox{.}\mkern2mu
 \raise4pt\hbox{.}\mkern2mu\raise7pt\hbox{.}\mkern1mu}} & \vdots \\
{{x^{{\beta ^k}}}}&{ \cdots \;}&{{x^{{\beta ^{k - 2}}}}}&{{x^{{\beta ^{k - 1}}}}}
\end{array}} \right) \hspace{0.2cm} \text{and}\hspace{0.2cm}
{H_2}(x) = \left( {\begin{array}{*{20}{c}}
{1 + {x^{ - \beta }}}& \cdots &{1 + {x^{ - {\beta ^{k - 1}}}}}&{1 + {x^{ - {\beta ^k}}}}\\
{1 + {x^{ - {\beta ^2}}}}& \cdots &{1 + {x^{ - {\beta ^k}}}}&{1 + {x^{ - {\beta ^1}}}}\\
 \vdots & {\mathinner{\mkern2mu\raise1pt\hbox{.}\mkern2mu
 \raise4pt\hbox{.}\mkern2mu\raise7pt\hbox{.}\mkern1mu}} & {\mathinner{\mkern2mu\raise1pt\hbox{.}\mkern2mu
 \raise4pt\hbox{.}\mkern2mu\raise7pt\hbox{.}\mkern1mu}} & \vdots \\
{1 + {x^{ - {\beta ^k}}}}&{ \cdots \;}&{1 + {x^{ - {\beta ^{k - 2}}}}}&{1 + {x^{ - {\beta ^{k - 1}}}}}
\end{array}} \right).
\end{align}
Then the first circulant array of $H_1(x)H_{2}(x^{-1})$ is
\begin{align}
\label{equ:QCSAequ1}
\renewcommand\arraycolsep{2pt}
\left[ {\begin{array}{*{20}{c}}
{\sum\limits_{j = 1}^k {\left( {{x^{{\beta ^j}}} + {x^{{\beta ^{2j}}}}} \right)} }&{\sum\limits_{j = 1}^k {{x^{{\beta ^j}}}\left( {1 + {x^{{\beta ^{(j + 1)(\bmod k)}}}}} \right)} }& \cdots &{\sum\limits_{j = 1}^k {{x^{{\beta ^j}}}\left( {1 + {x^{{\beta ^{(j + 1)(\bmod k)}}}}} \right)} }
\end{array}} \right],
\end{align}
and the other $k-1$ circulant arrays are cyclic shift of (\ref{equ:QCSAequ1}) to the right. Similarly, the first circulant array of $H_2(x)H_{1}(x^{-1})$ is
\begin{align}
\label{equ:QCSAequ2}
\renewcommand\arraycolsep{2pt}
\left[ {\begin{array}{*{20}{c}}
{\sum\limits_{j = 1}^k {\left( {{x^{ - {\beta ^j}}} + {x^{ - {\beta ^{2j}}}}} \right)} }&{\sum\limits_{j = 1}^k {{x^{ - {\beta ^j}}}\left( {1 + {x^{ - {\beta ^{(j + 1)(\bmod k)}}}}} \right)} }& \cdots &{\sum\limits_{j = 1}^k {{x^{ - {\beta ^j}}}\left( {1 + {x^{ - {\beta ^{(j + 1)(\bmod k)}}}}} \right)} }
\end{array}} \right],
\end{align}
and the other $k-1$ circulant arrays are cyclic shift of (\ref{equ:QCSAequ2}) to the right.
The result of $H_1(x)H_2(x^{-1})+H_2(x)H_1(x^{-1})$ is
\begin{equation}
\centering
\label{equ:45}
\renewcommand\arraycolsep{2pt}
\begin{array}{l}
\sum\limits_{j = 1}^k {\left( {{x^{ {\beta ^j}}} + {x^{{\beta ^{2j}}}}}\right)+\left({{x^{ - {\beta ^j}}} + {x^{ - {\beta ^{2j}}}}} \right)}\hspace{0.2cm} \text{on the diagonal,}\\
\sum\limits_{j = 1}^k {{x^{{\beta ^j}}}\left( {1 + {x^{{\beta ^{(j + 1)(\bmod k)}}}}} \right) + {x^{ - {\beta ^j}}}\left( {1 + {x^{ - {\beta ^{(j + 1)(\bmod k)}}}}} \right)} \hspace{0.2cm} \text{elsewhere.}
\end{array}
\end{equation}
As a consequence of \emph{Lemma \ref{lem:QRSproperty}}, if $2\in\mathcal{Q^R}$, then $2\beta^j\in\mathcal{Q^R}$ for $1\leq j\leq k$. On the other hand, if $2\in\mathcal{Q^{NR}}$, $2\beta^j\in\mathcal{Q^{NR}}$ for $1\leq j\leq k$. In both cases, the diagonal term shown in (\ref{equ:45}) generates entire $\mathcal{G}_{\mathbb{Z}_p}^{\times}$ twice. Further, as $\beta\in\mathcal{Q^R}$, $-\beta\in\mathcal{Q^{NR}}$, and both $\mathcal{Q^R}$ and $\mathcal{Q^{NR}}$ are cyclic, the off-diagonal term given in (\ref{equ:45}) also generates entire $\mathcal{G}_{\mathbb{Z}_p}^{\times}$ twice. Thus,
\begin{align}
{H_{1}}{H_{2}}^T + {H_{2}}{H_{1}}^T=2\left(\mathbb{I}_{proto}^{k \times k} \bigotimes {\mathbb{I}^{\mathcal{G}_{\mathbb{Z}_p}^{\times}}}\right) \equiv \mathbf{0}^{pk\times pk}\left(\bmod\hspace{0.1cm}2\right),
\end{align}
and $H_1$ and $H_2$ are commuting pairs.
\end{IEEEproof}

\begin{example}\label{exp:QCDSS} For $n=2$ and $p=7$, $\mathcal{Q^R} = \{\beta,\beta^2,\beta^3\}=\{2, 4, 1\}\left(\bmod 7\right)$. By \emph{Proposition \ref{prop:2}}, we obtain the following proto-matrix of $H$
\begin{equation*}
\centering
\label{equ:34}
\begin{array}{*{20}{l}}
\begin{array}{l}
{H_{proto}} = \left[ {{H_1}_{proto}|{H_2}_{proto} \boxplus \boldsymbol\emptyset_{proto}^{k \times k}} \right]\\
\renewcommand\arraystretch{0.9}
\renewcommand\arraycolsep{2pt}
 = \left[ {\begin{array}{c|c}
{\begin{array}{*{20}{c}}
{\left\{ {2,4,1} \right\}}\\
{2\cdot\left\{ {2,4,1} \right\}}\\
{4\cdot\left\{ {2,4,1} \right\}}
\end{array}}&{\begin{array}{*{20}{c}}
{\left\{ {5,3,6} \right\}}\\
{2\cdot\left\{ {5,3,6} \right\}}\\
{4\cdot\left\{ {5,3,6} \right\}}
\end{array}  \begin{array}{*{20}{c}}
{\boxplus\left( {0,0,0} \right)}\\
{\boxplus\left( {0,0,0} \right)}\\
{\boxplus\left( {0,0,0} \right)}
\end{array}}
\end{array}} \right]\\
 =
 \renewcommand\arraystretch{0.9}
\renewcommand\arraycolsep{4pt}
\left[ {\begin{array}{c|c}
{\begin{array}{*{20}{c}}
2&4&1\\
4&1&2\\
1&2&4
\end{array}}&{\begin{array}{*{20}{c}}
{0 \boxplus 5}&{0 \boxplus 3}&{0 \boxplus 6}\\
{0 \boxplus 3}&{0 \boxplus 6}&{0 \boxplus 5}\\
{0 \boxplus 6}&{0 \boxplus 5}&{0 \boxplus 3}
\end{array}}
\end{array}} \right].
\end{array}
\end{array}
\end{equation*}
The parity-check matrix $H$ is then obtained by lifting each element of $H_{proto}$ with CPM $P$ of size $7$, that is
\begin{equation}
\centering
\label{equ:QCexample}
H =
\renewcommand\arraystretch{1.0}
\renewcommand\arraycolsep{3pt}
\left[ {\begin{array}{c|c}
{\begin{array}{*{20}{c}}
{{P^{2}}}&{{P^{4}}}&{{P^{1}}}\\
{{P^{4}}}&{{P^{1}}}&{{P^{2}}}\\
{{P^{1}}}&{{P^{2}}}&{{P^{4}}}
\end{array}}&{\begin{array}{*{20}{c}}
{{P^{0}} + {P^{5}}}&{{P^{0}} + {P^{3}}}&{{P^{0}} + {P^{6}}}\\
{{P^{0}} + {P^{3}}}&{{P^{0}} + {P^{6}}}&{{P^{0}} + {P^{5}}}\\
{{P^{0}} + {P^{6}}}&{{P^{0}} + {P^{5}}}&{{P^{0}} + {P^{3}}}
\end{array}}
\end{array}} \right].
\end{equation}
$\Box$
\end{example}

\begin{proposition}
\label{prop:3}
For a positive integer $n$ and a prime $p=4n-1$, the parity-check matrix $H$ yields a $[[N,K,d_{min}]]=[[kp,kp-Rank(H),d_{min}]]$ QCS-A code, where $Rank(H)=k(p-1)+1$ when $n$ is odd and $Rank(H)=k(p-2)+1$ when $n$ is even.
$\blacksquare$
\end{proposition}
\begin{IEEEproof}
Let $\{d_1,d_2,\ldots,d_k\}$ be the $k$ elements of $\mathcal{Q^R}$, and $\sigma_1,\sigma_2,\ldots,\sigma_k$ be $k$ permutations of $\{1,2,\ldots, k\}$ such that $d_{\sigma_1(j)},d_{\sigma_2(j)},\ldots,d_{\sigma_k(j)}$ are distinct for every $1\leq j\leq k$. The parity-check matrix $H_1$ of Equation (\ref{equ:protoQCDSS}) of size $kp\times kp$ over $\mathbb{F}_2$ is then expressed as
\begin{equation}
H_1 =
\left[\begin{matrix}
I^{(d_{\sigma_1(1)})} & \cdots & I^{(d_{\sigma_1(k)})} \\
\vdots & \ddots & \vdots \\
I^{(d_{\sigma_k(1)})} & \cdots & I^{(d_{\sigma_k(k)})}
\end{matrix}\right]
= \left[\begin{matrix}
P^{d_{\sigma_1(1)}}_{11} & \cdots & P^{d_{\sigma_1(k)}}_{1k} \\
\vdots & \ddots & \vdots \\
P^{d_{\sigma_k(1)}}_{k1} & \cdots & P^{d_{\sigma_k(k)}}_{kk}
\end{matrix}\right],
\end{equation}
where each $P^{d}_{ij}$ denotes the $d$-th power of $P$.
Let $\alpha$ be a primitive $p$-th root of unity and $\mathbb{F}_2(\alpha)$ be the minimal finite field containing both $\mathbb{F}_2$ and $\alpha$. Denote by $V$ the $p\times p$ Vandermonde matrix generated by $\alpha$ over $\mathbb{F}_2$($\alpha$):
\begin{equation*}
V_p = \left[
\begin{matrix}
1 & 1 & \cdots & 1 \\
1 & \alpha & \cdots & \alpha^{p-1} \\
\vdots & \vdots & \cdots & \vdots \\
1 & \alpha^{p-1} & \cdots & \alpha^{(p-1)(p-1)}
\end{matrix}
\right].
\end{equation*}
Since $\alpha$ is a primitive $p$-th root of unity, $\alpha^i$ is a root of $x^p - 1$ and not equal to $1$ for all $1 \leq i \leq p-1$. Moreover, since $x^p - 1 = (x-1)(x^{p-1} + \cdots + 1)$, $\sum_{j=0}^{p-1}\alpha^{ij} = 0$ for all $1 \leq i \leq p-1$. Furthermore, since 2 and $p$ are co-prime, summation of $1$ by $p$ times is still equal to 1 over $\mathbb{F}_2$. It is then easy to check that the inverse of $V_p$ is:
\begin{align*}
V_p^{-1} = \left[
\begin{matrix}
1 & 1 & \cdots & 1 \\
1 & \alpha^{-1} & \cdots & \alpha^{-(p-1)} \\
\vdots & \vdots & \cdots & \vdots \\
1 & \alpha^{-(p-1)} & \cdots & \alpha^{-(p-1)(p-1)}
\end{matrix}
\right].
\end{align*}
For $0 \leq i \leq p-1$, denote by $\mathcal{D}(\alpha^i)$ the $p\times p$ matrix with diagonal entries equal to $\{1, \alpha^i, \cdots, \alpha^{i(p-1)}\}$, that is,
\begin{align*}
\mathcal{D}(\alpha^i) = \left[\begin{matrix}
1 & 0 & \cdots & 0 \\
0 & \alpha^i & \ddots & \vdots \\
\vdots & \ddots & \ddots & 0 \\
0 & \cdots & 0 & \alpha^{i(p-1)}
\end{matrix}\right].
\end{align*}
Thus,
\begin{align*}
I^{(1)} = P^1 = V_p\cdot \mathcal{D}(\alpha)\cdot V_p^{-1},
\end{align*}
and hence for any $0 \leq d \leq p-1$,
\[
I^{(d)} = P^d = V_p\cdot \mathcal{D}(\alpha^d) \cdot V_p^{-1}.
\]
The matrix $H_1$ can then be decomposed into
\begin{align}
H_1 = \left[\begin{matrix}
V_p & \mathbf{0} & \mathbf{0}  \\
\mathbf{0} & \ddots & \mathbf{0} \\
\mathbf{0} & \mathbf{0} & V_p
\end{matrix}\right] \cdot
\left[\begin{matrix}
\mathcal{D}(\alpha^{d_{\sigma_1(1)}}) & \cdots & \mathcal{D}(\alpha^{d_{\sigma_1(k)}})  \\
\vdots & \ddots & \vdots \\
\mathcal{D}(\alpha^{d_{\sigma_k(1)}}) & \cdots & \mathcal{D}(\alpha^{d_{\sigma_k(k)}})
\end{matrix}\right] \cdot
\left[\begin{matrix}
V_p^{-1} & \mathbf{0} & \mathbf{0}  \\
\mathbf{0} & \ddots & \mathbf{0} \\
\mathbf{0} & \mathbf{0} & V_p^{-1}
\end{matrix}\right].
\end{align}
{\color{black}
Write
\begin{align*}
\tilde{H_1} = \left[
\begin{matrix}
\mathcal{D}(\alpha^{d_{\sigma_1(1)}}) & \cdots & \mathcal{D}(\alpha^{d_{\sigma_1(k)}})  \\
\vdots & \ddots & \vdots \\
\mathcal{D}(\alpha^{d_{\sigma_k(1)}}) & \cdots & \mathcal{D}(\alpha^{d_{\sigma_k(k)}})
\end{matrix}
\right].
\end{align*}
Since both $\left[\begin{smallmatrix}
V_p & \mathbf{0} & \mathbf{0}  \\
\mathbf{0} & \ddots & \mathbf{0} \\
\mathbf{0} & \mathbf{0} & V_p
\end{smallmatrix}\right]$ and $\left[\begin{smallmatrix}
V_p^{-1} & \mathbf{0} & \mathbf{0}  \\
\mathbf{0} & \ddots & \mathbf{0} \\
\mathbf{0} & \mathbf{0} & V_p^{-1}
\end{smallmatrix}\right]$ have full rank $kp$, \[Rank(H_1) = Rank(\tilde{H_1}).\]

Note that $\tilde{H_1}$ can be regarded as a block matrix with $k \times k$ blocks each of which is a $p \times p$ diagonal matrix. We can then rearrange the columns and rows in $\tilde{H_1}$ to form a block diagonal matrix $\tilde{\tilde{H_1}}$ with each block of size $k\times k$, that is
\[
\tilde{\tilde{H_1}} = \left[\begin{matrix}
C_0 & \mathbf{0} & \cdots & \mathbf{0} \\
\mathbf{0} & C_1 & \ddots & \mathbf{0} \\
\vdots & \ddots & \ddots & \mathbf{0}\\
\mathbf{0} & \cdots & \mathbf{0} & C_{p-1}
\end{matrix}\right],
\]
where
$C_j =
\left[\begin{smallmatrix}
\alpha^{jd_{\sigma_1(1)}} & \cdots & \alpha^{jd_{\sigma_1(k)}} \\
\vdots & \vdots & \vdots \\
\alpha^{jd_{\sigma_{k}(1)}} & \cdots & \alpha^{jd_{\sigma_{k}(k)}}
\end{smallmatrix}\right]$
for all $0 \leq j \leq p-1$.
Obviously, \[Rank(\tilde{H_1}) = Rank(\tilde{\tilde{H_1}}) = Rank(C_0) + \cdots + Rank(C_{p-1}).\]
Since $C_0 = \left[ \begin{smallmatrix} 1 & \cdots & 1 \\ \vdots & \ddots & \vdots \\ 1 & \cdots & 1 \end{smallmatrix} \right]$, $Rank(C_0) = 1$.
Furthermore, since each row of $H_{1proto}$ is a cyclic shift of the first row, we denote
\begin{eqnarray*}
\sigma_2(1, \cdots, k) &=& (\sigma_1(k),\sigma_1(1),\cdots,\sigma_1(k-1)), \\
&\vdots \\
\sigma_k(1, \cdots, k) &=& (\sigma_1(2),\cdots,\sigma_1(k),\sigma_1(1)).
\end{eqnarray*}
Thus, $C_j$ is a circulant matrix over $\mathbb{F}_2$$(\alpha)$.
Let $U$ be the $k \times k$ cyclic permutation matrix in the same form as $P$ except for the different size. Then,
\begin{equation*}
C_j = \alpha^{jd_{\sigma_1(1)}}U^0 + \alpha^{jd_{\sigma_1(2)}}U^1 + \cdots + \alpha^{jd_{\sigma_1(k)}}U^{k-1}.
\end{equation*}
Let $\beta$ be a primitive $k^{th}$ root of unity, $\mathcal{D}(\beta^i)$ be the $k\times k$ diagonal matrix with diagonal entries equal to $\{1, \beta^i, \cdots, \beta^{i(k-1)}\}$, and $V_k$ be the $k\times k$ Vandermonde matrix (over $\mathbb{F}_2$$(\alpha)(\beta)$) generated by $\beta$. Since $U = V_k\cdot \mathcal{D}(\beta) \cdot V_k^{-1}$ and $U^i = V_k\cdot \mathcal{D}(\beta^i) \cdot V_k^{-1}$, we have
\[
C_j = V_k\cdot \left(\sum_{i = 0}^{k-1} \alpha^{jd_{\sigma_1(i+1)}}\mathcal{D}(\beta^i)\right) \cdot V_k^{-1}.
\]
Thus, the rank of $C_j$ is equal to the number of nonzero diagonal entries in $\sum_{i = 0}^{k-1} \alpha^{jd_{\sigma_1(i+1)}}\mathcal{D}(\beta^i)$. Equivalently, it is equal to $k - z$, where $z$ is the number of
roots of the polynomial $f_j(x) = \sum_{i = 0}^{k-1} \alpha^{jd_{\sigma_1(i+1)}}x^i$ that belong to $\{1, \beta, \cdots, \beta^{k-1}\}$.
{\color{black}
Since $p$ and $k$ are co-prime and $\alpha^p = \beta^k = 1$, it can be deduced that $f_j(\beta^i) \neq 0$ for all $1 \leq j \leq p-1$ and $1 \leq i \leq k-1$.
}
Next, note that $f_j(1) = \sum_{i = 0}^{k-1} \alpha^{jd_{\sigma_1(i+1)}} = \sum_{i = 1}^k \alpha^{jd_i}$. As a consequence of \emph{Lemma \ref{lemma:three}} and \emph{Corollary \ref{coro:T1Nodd}}: (i) when $n$ is odd, $\sum_{i = 1}^k \alpha^{jd_i} \neq 0$ for all $1 \leq j \leq p-1$; (ii) when $n$ is even, there are exactly $\frac{p-1}{2} = k$ elements in $\{\alpha, \cdots, \alpha^{p-1}\}$ which are roots of $\sum_{i = 1}^k x^{d_i}$. In all,
\begin{eqnarray*}
Rank(H_1) &=& Rank(C_0) + \cdots + Rank(C_{p-1}) \\
&=& \left\{\begin{matrix} 1+k(p-1), & \mathrm{when}~n~\mathrm{is~odd} \\ 1+k(p-1)-k, & \mathrm{when}~n~\mathrm{is~even} \end{matrix}\right..
\end{eqnarray*}}
Similarly, the rank of $H_2$ can be proved to be $1$ deficient from the rank of $H_1$. Since each $P_{ij}^{d}$ of $H_2$ has weight equal to $2$, \emph{i.e.,} $(I^{(d)}\boxplus I^{(0)})$, the diagonal matrix $\mathcal{D}$  will have the form
\begin{equation*}
\mathcal{D}(1 + \alpha^i) = \left[\begin{matrix}
0 & 0 & \cdots & 0 \\
0 & 1+\alpha^i & \ddots & \vdots \\
\vdots & \ddots & \ddots & 0 \\
0 & \cdots & 0 & 1+\alpha^{i(p-1)}
\end{matrix}\right].
\end{equation*}
Then $\tilde{H_2}$ will be written as
\begin{equation*}
\tilde{H_2} = \left[
\begin{matrix}
\mathcal{D}(1+\alpha^{d_{\sigma_1(1)}}) & \cdots & \mathcal{D}(1+\alpha^{d_{\sigma_1(k)}})  \\
\vdots & \ddots & \vdots \\
\mathcal{D}(1+\alpha^{d_{\sigma_k(1)}}) & \cdots & \mathcal{D}(1+\alpha^{d_{\sigma_k(k)}})
\end{matrix}
\right].
\end{equation*}
Note that the first entry of $\mathcal{D}(1+\alpha^i)$ is always $0$. By rearranging the columns and rows of $\tilde{H_2}$, we turn $\tilde{H_2}$ into a block diagonal matrix $\tilde{\tilde{H_2}}$ of same format as $\tilde{\tilde{H_1}}$ with $Rank(C_{0})=0$. Hence, $Rank(H_2) = Rank(H_1)-1$.
Furthermore, we know that the row rank of a matrix equal to its column rank. We also know that there exists a sub-matrix $H^{sub}$ with non-zero determinant of size $Rank(H_1)\times Rank(H_1)$. Therefore, the rank of the parity-check matrix $H$ is given by $Rank(H)=max\left\{Rank(H_1), Rank(H_2)\right\} = Rank(H_1)$.
\end{IEEEproof}

\begin{remark}
\label{remark:1}
If $n$ is odd and $k$ is divisible by $3$, then $Rank(H_1)=Rank(H_2) +1 = k(p-3) + 1$.
$\blacksquare$
\end{remark}

We now give the following method to construct QCS-A codes with the maximum number of independent stabilizer generators.

\textbf{\emph{Construction of QCS-A Codes:}}
For $n\in\mathbb{Z}^{+}$ and a prime $p=4n-1$, a parity-check matrix $H=[H_1|H_2]$ can be lifted from a proto-matrix  $H_{proto}=[H_{1proto}|H_{2proto}]$ that is constructed from $\mathcal{Q^R}$ and $\mathcal{Q^{NR}}$. We obtain the QCS-A code $H^{sub}=[H_1^{sub}|H_2^{sub}]$ based on the following procedures:

\begin{enumerate}
\item For an odd $n$, if $3$ is a divisor of $k$ (\emph{i.e.,} non-prime), we first remove one arbitrary circulant array from $k$ arrays of $H$. We then choose an additional $k-2$ arbitrary arrays from the rest $k-1$ circulant arrays of $H$. Finally, we remove an arbitrary row from each of the chosen circulant arrays.
\item For an odd $n$, if $k$ is a prime, or $k$ is a non-prime that is not divisible by $3$, we choose arbitrary $k-1$ circulant arrays out of $k$ arrays of $H$. Then we remove an arbitrary row from each of the chosen circulant arrays.
\item For an even $n$, we first choose arbitrary $k-1$ circulant arrays out of $k$ arrays of $H$, and we remove an arbitrary row from each of the chosen circulant arrays. Then we remove an additional row from each one of the $k$ circulant arrays.

\end{enumerate}
The resulting parity-check matrix $H^{sub} = [H_1^{sub}|H_{2}^{sub}]$ is a $[[pk, pk-Rank(H),d_{min}]]$ QCS-A code.
$\Box$

Note that the all-zero proto-matrix $\boldsymbol{\emptyset}^{k\times k}_{proto}$ presented in \emph{Proposition \ref{prop:2}} can also be adjunct to $H_{1proto}$. The properties shown in \emph{Propositions \ref{prop:2}} and \emph{\ref{prop:3}} are similar.

\begin{figure*}
\centering
\includegraphics[width=5.5in]{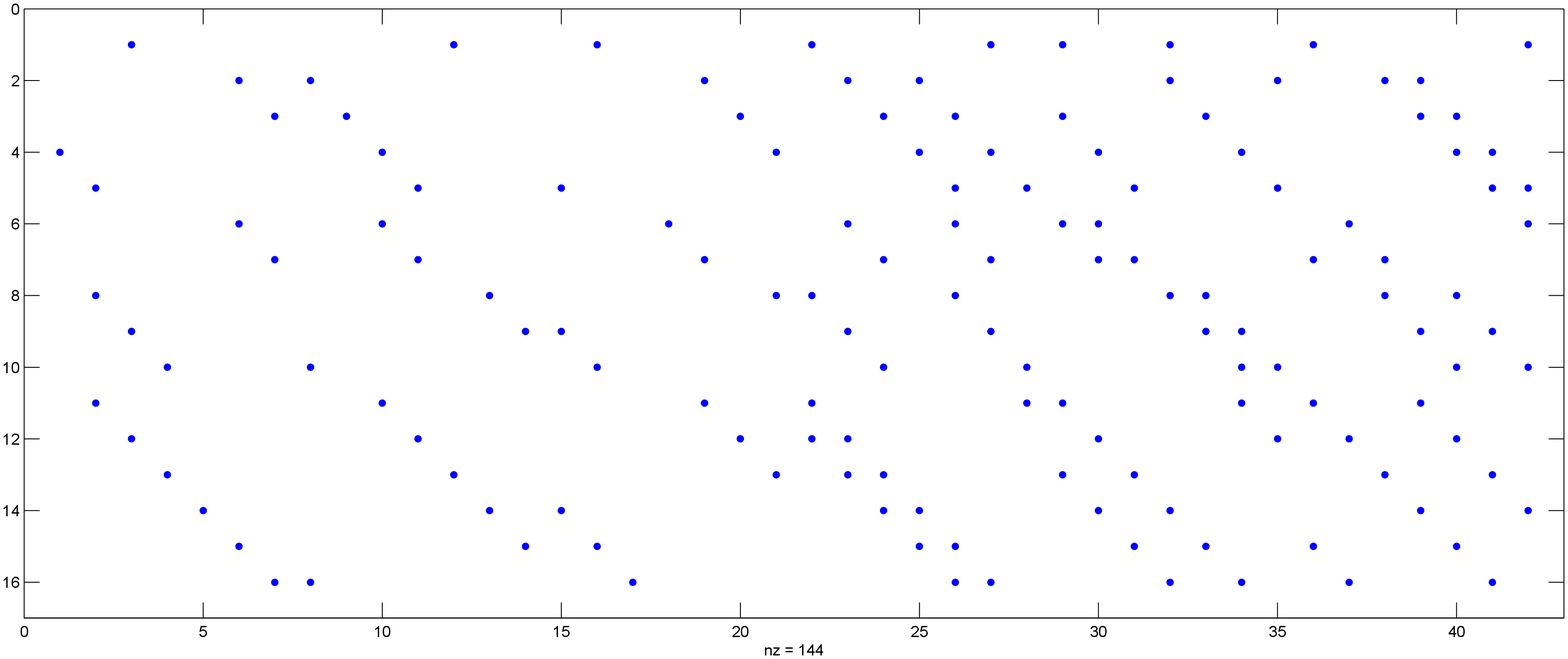}
\caption{\hspace{0.1cm}$[[21, 5, 4]]$ QCS-A code in \emph{Example \ref{exp:QCDSS}} and \emph{\ref{exp:5}. }}
\label{fig:ExampleCode}
\end{figure*}

\begin{example}\label{exp:5} Continue from \emph{Example} \ref{exp:QCDSS}.
Since $n$ is even and $k$ is a prime, by using the third construction method for QCS-A codes, we remove arbitrary $k-1=2$ rows from arbitrary $k-1=2$ circulant arrays plus additional one row from  each circulant array. The parity-check matrix $H$ of this example is depicted in Fig. \ref{fig:ExampleCode}, where the rows $\{2, 3, 8, 11, 21\}$ are removed from the lifted parity-check matrix. The resulting number of linearly independent stabilizer generators is $16$, and thus we can encode $K=5$ logical qubits into $N=21$ physical qubits. The minimum distance of this code is $d_{min}=4$ and it is a $[[21,5,4]]$ QCS-A code. $\Box$
\end{example}

%

\subsection{QCS-B Codes from QR set of size p = 4n+1}
We now give another type of QCS codes, Type-II QCS-B codes, designed from quadratic residue set of size $p=4n+1$. In this case, no adjunct of element $0$ is required.
Let $h_P(\mathcal{Q^R}) = \left[\mathcal{Q^R}(1) \mathcal{Q^R}(2) \ldots,\mathcal{Q^R}(k)\right]$ $=\left[\beta,\beta^2,\ldots,\beta^k\right]$, the same format given in Equation (\ref{equ:30}).
From \emph{Lemma \ref{lem:QRSproperty}}, since $\alpha^{2i-1}\cdot \alpha^{2i}\in\mathcal{Q^{NR}}$ for $1\leq i\leq \frac{p-1}{2}$, where $\alpha^{2i}\in\mathcal{Q^R}$ and $\alpha^{2i-1}\in\mathcal{Q^{NR}}$, we position $h_P(\mathcal{Q^{NR}})$ in the following format
\begin{align}
\label{equ:QCDSS2}
h_P(\mathcal{Q^{NR}}) = min(\mathcal{Q^{NR}})h_P(\mathcal{Q^{R}})\hspace{0.1cm} \left(\bmod \hspace{0.1cm}p\right).
\end{align}
Unlike the method used for QCS-A codes, we design $H_{1proto}$ by reversing the direction of cyclic shifts of $h_P(\mathcal{Q^{R}})$.
We denote the $i$-th cyclic right shift of $h_P(\mathcal{Q^{R}})$ as
\begin{align}
\centering
\renewcommand\arraystretch{1.0}
\renewcommand\arraycolsep{3pt}
\label{equ:QCSB}
\begin{array}{l}
circ\left(h_P(\mathcal{Q^R})\right)^{-i}\equiv
{h_P}\left( {{\beta ^{-i}}\mathcal{Q^R}} \right) = \left[
{{\beta ^{-i}}\mathcal{Q^R}( 1 )}\hspace{0.1cm}{{\beta ^{-i}}\mathcal{Q^R}( 2 )}\hspace{0.1cm} \cdots\hspace{0.1cm} {{\beta ^{-i}}\mathcal{Q^R}( k )}\right].\\
\end{array}
\end{align}
By juxtaposing different shifts of ${h_P}\left( {{\beta ^{-i}}\mathcal{Q^R}} \right)$, we obtain
\begin{align}
\renewcommand\arraystretch{0.8}
\renewcommand\arraycolsep{3pt}
\label{equ:QCDSS2b}
{H_{1proto}} = \left[ \begin{array}{l}
{h_P}({\mathcal{Q^{R}}})\\
circ{({h_P}({\mathcal{Q^{R}}}))^{ - 1}}\\
 \vdots \\
circ{({h_P}({\mathcal{Q^{R}}}))^{ - k + 1}}
\end{array} \right].
\end{align}
Furthermore, based on (\ref{equ:QCDSS2}), we construct $H_{2proto}$ as
\begin{align}
\renewcommand\arraystretch{0.9}
\renewcommand\arraycolsep{2.5pt}
\label{equ:QCDSS2c}
{H_{2proto}} = \left[ {\begin{array}{*{20}{l}}
{{h_P}({\mathcal{Q^{NR}}})}\\
{{\beta }{h_P}({\mathcal{Q^{NR}}})}\\
{\beta ^2}{h_P}({\mathcal{Q^{NR}}})\\
 \vdots \\
{{\beta ^{k - 1}}{h_P}({\mathcal{Q^{NR}}})}
\end{array}} \right] \equiv \left[ {\begin{array}{*{20}{l}}
{min ({\mathcal{Q^{NR}}}){h_P}({\mathcal{Q^R}})}\\
{min ({\mathcal{Q^{NR}}})circ{{({h_P}({\mathcal{Q^R}}))}^1}}\\
min ({\mathcal{Q^{NR}}})circ{({h_P}({\mathcal{Q^R}}))^2}\\
 \vdots \\
{min ({\mathcal{Q^{NR}}})circ{{({h_P}({\mathcal{Q^R}}))}^{k - 1}}}
\end{array}} \right].
\end{align}
The constructed proto-matrices $H_{1proto}$ and $H_{2proto}$ are square matrices of $k$ different permutations of $\mathcal{Q^R}$ and $\mathcal{Q^{NR}}$, respectively. They are also Latin squares of order $k$ since every element of $\mathcal{Q^R}$ (resp. $\mathcal{Q^{NR}}$) appears exactly once in every row and every column. However, only $H_{2proto}$ is a commutative Latin square, whereas $H_{1proto}\neq H_{1proto}^T$ is not.

\begin{proposition}
\label{prop:QCDSS2}
For a positive integer $n$ and $p = 4n+1$, let $h_P(\mathcal{Q^R})=\left[\mathcal{Q^R}(1),\mathcal{Q^R}(2),\ldots,\mathcal{Q^R}(k)\right]$ and $h_P(\mathcal{Q^{NR}})=min(\mathcal{Q^{NR}})h_P(\mathcal{Q^{R}})(\hspace{-0.13cm}\bmod \hspace{-0.07cm}p)$. The parity-check matrix $H$ lifted from $H_{proto}=[H_{1proto}|H_{2proto}]$ as specified in (\ref{equ:QCDSS2b}) and (\ref{equ:QCDSS2c}) satisfies the SIP constraint which yields a
$[[N,K,d_{min}]]=[[pk,pk-Rank(H),d_{min}]]$ QCS-B code with $Rank(H)=k(p-1)+1$ when $n$ is odd and $Rank(H)=k(p-2)+1$ when $n$ is even.
\end{proposition}
\begin{IEEEproof}
Let $\gamma=min(\mathcal{Q^{NR}})$, we represent $H_1$ and $H_2$ in polynomial form of circulant arrays
\begin{align}
\setlength{\arraycolsep}{1.5pt}
\renewcommand{\arraystretch}{0.6}
{H_1(x)} = \left( {\begin{array}{*{20}{c}}
{{x^\beta }}&{{x^{{\beta ^2}}}}& \cdots &{{x^{{\beta ^k}}}}\\
{{x^{{\beta ^k}}}}&{{x^\beta }}& \cdots &{{x^{{\beta ^{k - 1}}}}}\\
 \vdots & \ddots & \ddots & \vdots \\
{{x^{{\beta ^2}}}}&{{x^{{\beta ^3}}}}& \cdots &{{x^{{\beta }}}}
\end{array}} \right)
\hspace{0.2cm}\text{and}\hspace{0.2cm}
{H_2(x)} = \left( {\begin{array}{*{20}{c}}
{{x^{\gamma \beta }}}&\cdots&{{x^{\gamma \beta^{k-1} }}} &{{x^{\gamma {\beta ^k}}}}\\
{{x^{\gamma {\beta ^2}}}}&\cdots& {{x^{\gamma \beta^k }}} &{{x^{\gamma {\beta}}}}\\
 \vdots & \iddots & \iddots & \vdots \\
{{x^{\gamma {\beta ^k}}}}& \cdots &{{x^{\gamma {\beta ^{k-2}}}}}&{{x^{\gamma {\beta ^{k - 1}}}}}
\end{array}} \right).
\end{align}
Then the first circulant array of $H_1(x)H_2(x^{-1})$ is
\begin{align}
\label{equ:QSC2proofa}
\left[ {\begin{array}{*{20}{c}}
{\sum\limits_{j = 1}^k {{x^{{\beta ^j}\left( {1 - \gamma {\beta ^0}} \right)}}} }&{\sum\limits_{j = 1}^k {{x^{{\beta ^j}\left( {1 - \gamma {\beta ^1}} \right)}}} }& \cdots &{\sum\limits_{j = 1}^k {{x^{{\beta ^j}\left( {1 - \gamma {\beta ^{k - 1}}} \right)}}} }
\end{array}} \right],
\end{align}
and the other $k-1$ circulant arrays are cyclic shift of Equation (\ref{equ:QSC2proofa}) to the left. Similarly, the first circulant array of $H_2(x)H_1(x^{-1})$ is
\begin{align}
\label{equ:QSC2proofb}
\left[ {\begin{array}{*{20}{c}}
{\sum\limits_{j = 1}^k {{x^{{\beta ^j}\left( {\gamma {\beta ^0} - 1} \right)}}} }&{\sum\limits_{j = 1}^k {{x^{{\beta ^j}\left( {\gamma {\beta ^1} - 1} \right)}}} }& \cdots &{\sum\limits_{j = 1}^k {{x^{{\beta ^j}\left( {\gamma {\beta ^{k - 1}} - 1} \right)}}} }
\end{array}} \right],
\end{align}
and the other $k-1$ circulant arrays are cyclic shift of Equation (\ref{equ:QSC2proofb}) to the left.
Note that $\gamma-1\equiv -(1-\gamma) (\bmod \hspace{0.1cm}p)$. We obtain the first circulant array of ${H_1}\left( x \right){H_2}\left( {{x^{ - 1}}} \right) + {H_2}\left( x \right){H_1}\left( {{x^{ - 1}}} \right)$ containing only the term
\begin{align}
\label{equ:QSC2proofc}
{{x^{\left( {1 - \gamma {\beta ^i}} \right)}}\sum\limits_{j = 1}^k {\left( {{x^{{\beta ^j}}} + {x^{ - {\beta ^j}}}} \right)} }
\end{align}
for $0 \le i \le k - 1$. Since for a prime $p=4n+1$, both $\{\beta,-\beta\}\in\mathcal{Q^R}$. This implies that $\sum\limits_{j = 1}^k {\left( {{x^{{\beta ^j}}} + {x^{ - {\beta ^j}}}} \right)}\equiv 0$. Hence, $H_1(x)$ and $H_2(x)$ are commuting pairs.
Moreover, using the similar way in the proof of \emph{Proposition \ref{prop:3}}, the rank of $H$ can be shown to be either $Rank(H_1)=Rank(H_2)=k(p-1)+1$ for $n$ is odd or $Rank(H_1)=Rank(H_2)=k(p-2)+1$ for $n$ is even. Thus, $Rank(H)=Rank(H_1)=Rank(H_2)$.
\end{IEEEproof}

The methods $2)$ and $3)$ of the \emph{Construction of QCS-A Codes} can be applied here to generate a QCS-B code.
We now see an example.
\begin{example}
\label{exp:QCSII}
For $n=3$ and $p=13$, we have $\mathcal{Q^R}=\{4, 3, 12, 9, 10, 1\}$ and $\mathcal{Q^{NR}}=\{2, 5, 6, 7, 8, 11\}$ with $k=|\mathcal{Q^R}|=|\mathcal{Q^{NR}}|=\frac{p-1}{2}=6$. Let $\beta=4$ and $\gamma=min\left(\mathcal{Q^{NR}}\right)=2$. Then
\begin{align*}
\begin{array}{l}
{H_{proto}} = \left[ {{H_{1proto}}|{H_{2proto}}} \right] = \\
\setlength{\arraycolsep}{1.5pt}
\renewcommand{\arraystretch}{0.6}
\left[ {\begin{array}{c|c}
{\begin{matrix}
4&3&{12}&9&{10}&1\\
1&4&3&{12}&9&{10}\\
{10}&1&4&3&{12}&9\\
9&{10}&1&4&3&{12}\\
{12}&9&{10}&1&4&3\\
3&{12}&9&{10}&1&4
\end{matrix}}&{2\left( {\begin{matrix}
4&3&{12}&9&{10}&1\\
3&{12}&9&{10}&1&4\\
{12}&9&{10}&1&4&3\\
9&{10}&1&4&3&{12}\\
{10}&1&4&3&{12}&9\\
1&4&3&{12}&9&{10}
\end{matrix}} \right)}
\end{array}} \right]
=\left[ {\begin{array}{c|c}
{\begin{matrix}
4&3&{12}&9&{10}&1\\
1&4&3&{12}&9&{10}\\
{10}&1&4&3&{12}&9\\
9&{10}&1&4&3&{12}\\
{12}&9&{10}&1&4&3\\
3&{12}&9&{10}&1&4
\end{matrix}}&{ {\begin{matrix}
8&6&{11}&5&{7}&2\\
6&{11}&5&{7}&2&8\\
{11}&5&{7}&2&8&6\\
5&{7}&2&8&6&{11}\\
{7}&2&8&6&{11}&5\\
2&8&6&{11}&5&{7}
\end{matrix}}}
\end{array}} \right]
\end{array}.
\end{align*}
By lifting each element of $H_{proto}$ with CPM $P$ of size $13$, we obtain
\begin{align*}
\setlength{\arraycolsep}{2.0pt}
\renewcommand{\arraystretch}{0.7}
H = \left[ {\begin{array}{c|c}
{\begin{matrix}
{{P^4}}&{{P^3}}&{{P^{12}}}&{{P^9}}&{{P^{10}}}&{{P^1}}\\
{{P^1}}&{{P^4}}&{{P^3}}&{{P^{12}}}&{{P^9}}&{{P^{10}}}\\
{{P^{10}}}&{{P^1}}&{{P^4}}&{{P^3}}&{{P^{12}}}&{{P^9}}\\
{{P^9}}&{{P^{10}}}&{{P^1}}&{{P^4}}&{{P^3}}&{{P^{12}}}\\
{{P^{12}}}&{{P^9}}&{{P^{10}}}&{{P^1}}&{{P^4}}&{{P^3}}\\
{{P^3}}&{{P^{12}}}&{{P^9}}&{{P^{10}}}&{{P^1}}&{{P^4}}
\end{matrix}}&{\begin{matrix}
{{P^8}}&{{P^6}}&{{P^{11}}}&{{P^5}}&{{P^7}}&{{P^2}}\\
{{P^6}}&{{P^{11}}}&{{P^5}}&{{P^7}}&{{P^2}}&{{P^8}}\\
{{P^{11}}}&{{P^5}}&{{P^7}}&{{P^2}}&{{P^8}}&{{P^6}}\\
{{P^5}}&{{P^7}}&{{P^2}}&{{P^8}}&{{P^6}}&{{P^{11}}}\\
{{P^7}}&{{P^2}}&{{P^8}}&{{P^6}}&{{P^{11}}}&{{P^5}}\\
{{P^2}}&{{P^8}}&{{P^6}}&{{P^{11}}}&{{P^5}}&{{P^7}}
\end{matrix}}
\end{array}} \right].
\end{align*}
It can be checked that H satisfies the SIP constraint.
Furthermore, since $Rank(H)=6(13-1)+1=73$, we remove $5$ rows separately from $5$ different circulant arrays. Thus, we obtain a $[[73, 5, d_{min}]]$ QCS-B code.
$\Box$
\end{example}

%

\section{Constructed Codes and Simulation Results}
We constructed Type-I stabilizer codes of length $N=4n+1$ and $N=4n-1$ for $n\leq 25$ and the results are listed in TABLEs \ref{tab:TypeIa} and \ref{tab:TypeIb}. The codes in TABLE \ref{tab:TypeIa} are Type-I $[[N,K,d_{min}]]=[[4n+1,1,d^{\dag}\geq d_{min}\geq 3]]$ stabilizer codes. The corresponding $d^{\dag}$ of the codes, which denotes the minimum weight of an operator $E\in\mathcal{M}$, is also shown in the table. From these two tables, it can be seen that $d_{min}<d^{\dag}$ for all $n$, which means they are all non-degenerate stabilizer codes. Further, we find that for the code lengths $N=4n+1$, $n=1, 3$ and $7$, our constructed Type-I stabilizer codes in TABLE \ref{tab:TypeIa} achieve the highest minimum distance as given in \cite{Grassl}. 
For $n=9$, the constructed $[[37,1,12]]$ Type-I stabilizer code satisfies the distance bound $11\leq d_{min}\leq 13$ given in \cite{Grassl}. Moreover, our constructed Type-I stabilizer codes, $[[53,1,15]]$, $[[61,1,17]]$ and $[[101,1,21]]$ meet the lower bound of the achievable minimum distance given in \cite{Grassl}. As shown in TABLE \ref{tab:TypeIa}, for $n=1$, our Type-I stabilizer code is equivalent to the \emph{perfect} $[[5,1,3]]$ code \cite{laf1996}, and it has $d^{\dag}=4>d_{min}$ shown in brackets, where $d^{\dag} = 4n$. Also, for $n=3$ and $7$, our $[[13,1,5]]$ and $[[29,1,11]]$ Type-I stabilizer codes are equivalent to the codes proposed in \cite{CalderRainShorSloane1997}.
Furthermore, the codes in TABLE \ref{tab:TypeIb} are Type-I $[[N,K,d_{min}]]=[[4n-1,2n-1,2]]$ codes, where the code rate $\frac{K}{N}$ is approximately half and $d_{min}=2$ for any even $n$ that gives a prime $p=4n-1$. The minimum distance $d^{\dag}$ of stabilizer $\mathcal{S}$ is also listed in the table.

In TABLE \ref{tab:QCS}, we show the constructed some Type-II QCS codes of length $N=kp$ and $p=4n\pm 1$, where $n$ is both even and odd. We give the exact minimum distance for the constructed codes of length $N=10$ and $21$. The distance range for the constructed QCS codes of length $N=21$ represents different $[[21, K, d_{min}]]$ QCS codes constructed with same $K$ and different $d_{min}$.


Note that the proposed Type-II QCS-A codes of length $N=kp$ exist in the form of $H_{proto}=[H_{1proto}|H'_{2proto}=H_{2proto}\boxplus \emptyset^{k\times k}_{proto}]$. We also know that the proposed method enables $\emptyset^{k\times k}_{proto}$ adjunct to either $H_{1proto}$ or $H_{2proto}$. This implies that there are four different arrangements for $H_{proto}$, \emph{i.e.,} $H_{proto}=\left[H'_{2proto}|H_{1proto}\right]$, $H_{proto}=\left[H'_{1proto}|H_{2proto}\right]$, $H_{proto}=\left[H_{1proto}|H'_{2proto}\right]$ and
$H_{proto}=\left[H_{2proto}|H'_{1proto}\right]$. Interestingly, if we swap the position of $H_{1proto}$ and $H'_{2proto}$, we might obtain two different codes. More importantly, by removing different rows, different codes with different minimum distance might also be obtained. TABLE \ref{tab:QCS-A21} illustrates  various $[[21,5,d_{min}]]$ and $[[21,6,d_{min}]]$ QCS-A codes constructed using different arrangement of proto-matrix $H_{proto}$, and by removing different rows of the lifted parity-check matrix $H$.
Let $H_{proto}=[H_{1proto}|H'_{2proto}]$, from the table, for $[[21, 5, d_{min}]]$ QCS-A code, $d_{min}=4$ if rows $\{7, 11, 12, 14, 21\}$ are removed from the lifted parity-check matrix $H$. However, if $H_{proto}=[H'_{2proto}|H_{1proto}]$, we obtain a $[[21, 5, 5]]$ QCS-A code if the same rows $\{7, 11, 12, 14, 21\}$ are removed. Moreover, if $H_{proto}=\left[H'_{1proto}|H_{2proto}\right]$ and rows $\{5, 8, 9, 13, 21\}$ are removed, we obtain a $[[21, 5, 3]]$ QCS-A code. Furthermore, for $[[21,6,d_{min}]]$ QCS-A codes, we have $d_{min}=4$ (resp. $d_{min}=3$) if $H_{proto}=[H_{1proto}|H'_{2proto}]$ (resp. $H_{proto}=[H'_{2proto}|H_{1proto}]$) and the rows $\{3, 7, 11, 12, 14, 21\}$ are removed. Similarly, we can also construct $[[21, 6, 4]]$ and $[[21,6, 2]]$ QCS-A codes when rows $\{4, 7, 11, 12, 14, 21\}$ are removed. If $\emptyset^{k\times k}$ is adjunct to either $H_{1proto}$ or $H_{2proto}$ and the rows $\{7, 11, 12, 14, 15, 21\}$ are removed in both cases, we obtain a $[[21, 6, 4]]$ or $[[21, 6, 1]]$ QCS-A code, respectively.

Now we compare our constructed codes with the benchmarks of asymptotic code efficiency for quantum codes.
Recall that the quantum Hamming bound is \cite{Gottesman1996}
\begin{align}
\label{equ:qhb}
\sum_{j=0}^{t} 3^j{N \choose j}\leq 2^{m},
\end{align}
and the code efficiency is asymptotically upper bounded by
\begin{align}
\label{equ:QHBA}
\frac{K}{N}\leq 1 - \delta^Q \log_2(3) - h_2(\delta^Q),
\end{align}
where $m = N-K$ is the number of stabilizer generators, $t = \lfloor\frac{d_{min}-1}{2}\rfloor$ is the number of correctable errors, and $\delta^Q=\frac{t}{N}$. In (\ref{equ:QHBA}), $h_2(*)$ is the binary entropy function $h_2(x)=-x\log_2(x)-(1-x)\log_2(1-x)$.
The quantum Gilbert-Varshamov (GV) bound is \cite{CalderRainShorSloane1997}
\begin{align}
\label{equ:QGVB}
\sum_{j=0}^{2t} 3^j{N \choose j}\leq 2^{m},
\end{align}
and the code efficiency is asymptotically lower bounded by
\begin{align}
\label{equ:QGVBA}
\frac{K}{N}\geq 1 - 2\delta^Q \log_2(3) - h_2(2\delta^Q).
\end{align}
For an $[[N,K,d_{min}]]$ CSS code, we have the Gilbert-Varshamov bound \cite{Calderbank1996}
\begin{align}
\frac{K}{N}\geq 1 - 2h_2(2\delta^Q).
\end{align}
And any (degenerate and non-degenerate) quantum code must satisfy the quantum Singleton bound \cite{KnillLaf1997}
\begin{align}
\label{equ:QSB}
N - K\geq 4t.
\end{align}
Hence, the asymptotic code efficiency is given by
\begin{align}
\label{equ:QSBA}
\frac{K}{N}\leq 1- 4\delta^Q.
\end{align}

\begin{figure*}[ht]
\centering
\includegraphics[width=6.8in]{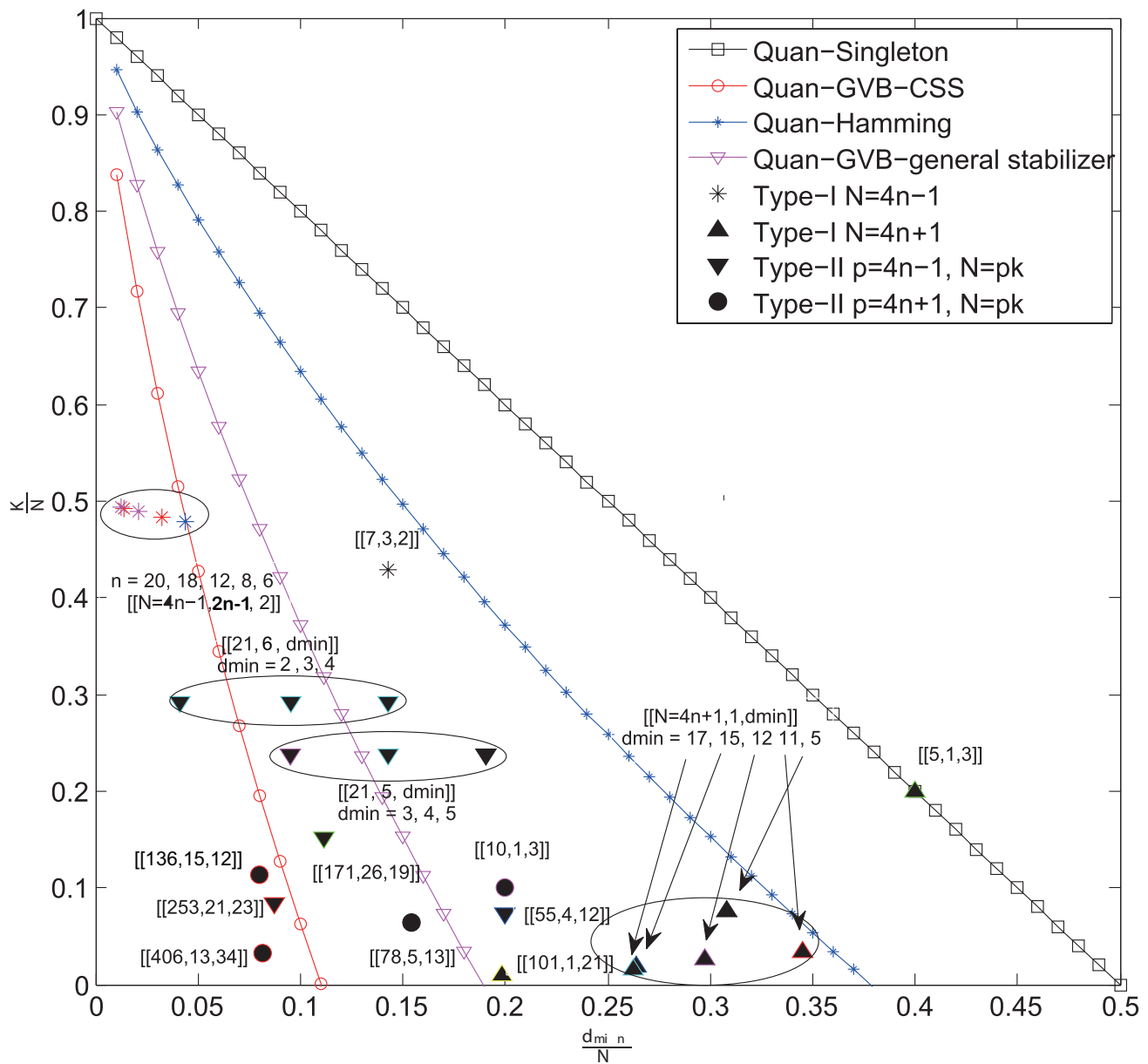}
\caption{Known asymptotic quantum coding bounds and some designed Type-I and Type-II stabilizer codes.}
\label{fig:quantumBounds}
\end{figure*}

The above known quantum bounds in the literature are depicted in Fig. \ref{fig:quantumBounds} with code rate $\frac{K}{N}$ in terms of its normalized distance $\frac{d_{min}}{N}$, where $\frac{d_{min}}{N}\approx 2\delta^Q$ for sufficiently large $N$. We also depicted our constructed codes in Fig. \ref{fig:quantumBounds} for comparison.

From the figure, the equality of the quantum Hamming bound (\ref{equ:qhb}) and the quantum Singleton bound (\ref{equ:QSB}) holds for $[[5,1,3]]$ Type-I stabilizer code when $n=1$. In this case, the number of correctable error $t$ is equal to $n$. For other Type-I stabilizer codes of $N=4n+1$ with $N>5$, the code efficiency is upper bounded by the quantum Hamming bound for $t<n$ or $d_{min}<2n+1$.
Furthermore, for small value of $n$, the constructed $[[21,5,4]]$, $[[21,5,5]]$, $[[21, 6, 4]]$ and $[[55,4,12]]$ Type-II QCS-A codes and $[[10,1,3]]$ Type-II QCS-B code satisfy the quantum GV bound for general stabilizer codes, and upper bounded by quantum Hamming bound.
On the other hand, for large value of $n$, weaker lower bound can be satisfied for the constructed QCS codes, \emph{e.g.,} the $[[78, 5, 13]]$ QCS-B code and $[[171, 26, 19]]$ QCS-A code satisfy the quantum GV bound for CSS codes.
\begin{table}[ht]
\renewcommand{\arraystretch}{1.5}
\caption{Type-I stabilizer codes of length $N=4n+ 1$ for $n\leq 25$. $d^{\dag}$ is the minimum weight of operator $E\in\mathcal{S}$. Underlined numbers indicate that $d_{min}$ meets the lower bound of the achievable minimum distance given in \cite{Grassl}. Number with brackets is the Perfect code in \cite{laf1996}.}
\centering
\begin{tabular}{c c c}
\hline\hline
\vspace{0.05cm}
$n$ & ${[[N,K,d_{min}]]}$ & $d^{\dag}\leq 2n$\\ \hline
$1$ &  [[5,1,3]] & (4)  \\ \hline
$3$ &  [[13,1,5]] & 6  \\ \hline
$7$ &  [[29,1,11]] & 12  \\ \hline
$9$ &  [[37,1, 12]]& 12  \\ \hline
$13$&  [[53,1,\underline{15}]]& 16   \\ \hline
$15$&  [[61,1,\underline{17}]]& 18   \\ \hline
$25$& [[101,1,\underline{21}]]& 22  \\
\hline\hline
\end{tabular}
\label{tab:TypeIa}
\end{table}

\begin{table}[ht]
\renewcommand{\arraystretch}{1.5}
\caption{Type-I stabilizer codes of length $N = 4n - 1$ for $n\leq 25$. $d^{\dag}$ is the minimum weight of operator $E\in\mathcal{S}$.}
\centering
\begin{tabular}{c c c}
\hline\hline
$n$ & $[[N,K,d_{min}]]$ & $d^{\dag}$ \\ \hline
$2$ & $[[7,3,2]]$ & $4$ \\ \hline
$6$ & $[[23,11,2]]$& $8$ \\ \hline
$8$ & $[[31,15,2]]$& $8$ \\ \hline
$12$ & $[[47,23,2]]$& $12$ \\ \hline
$18$ & $[[71,35,2]]$& $16$ \\ \hline
$20$ & $[[79,39,2]]$& $16$ \\
\hline\hline
\end{tabular}
\label{tab:TypeIb}
\end{table}

\begin{table}[ht]
\renewcommand{\arraystretch}{1.5}
\renewcommand{\tabcolsep}{2.0pt}
\caption{Type-II QCS codes of length $N=kp$ and $p=4n\pm1$.}
\centering
\begin{tabular}{c c c c c}
\hline\hline
\vspace{0.05cm}
$n$ & Type& $p=4n\pm1$ & $k=\frac{p-1}{2}$ & $[[N=kp,K,d_{min}]]$ \vspace{0.05cm} \\ \hline
$1$ & QCS-B  &$5$ & $2$ & $[[10, 1, 3]]$ \vspace{0.05cm}\\ \hline
$2$ & QCS-A  &$7$ & $3$ & $[[21, 5, 3-5]]$ \vspace{0.05cm}\\ \hline
$2$ & QCS-A  &$7$ & $3$ & $[[21, 6, 1-4]]$ \vspace{0.05cm}\\ \hline
$3$ & QCS-A  &$11$ & $5$ & $[[55, 4, \leq{12}]]$ \vspace{0.05cm} \\ \hline
$3$ & QCS-B  &$13$ & $6$ & $[[78, 5, \leq{13}]]$ \vspace{0.05cm}\\ \hline
$4$ & QCS-B  &$17$ & $8$ & $[[136, 15, \leq{12}]]$ \vspace{0.05cm}\\ \hline
$5$ & QCS-A  &$19$ & $9$ & $[[171, 26, \leq{19}]]$ \vspace{0.05cm} \\ \hline
$6$ & QCS-A  &$23$ & $11$& $[[253, 21, \leq{23} ]]$ \vspace{0.05cm} \\ \hline
$7$ & QCS-B  &$29$ & $14$& $[[406, 13, \leq{34}]]$ \vspace{0.05cm} \\
\hline\hline
\end{tabular}
\label{tab:QCS}
\end{table}

\begin{table}[ht]
\renewcommand{\arraystretch}{1.5}
\renewcommand{\tabcolsep}{2.0pt}
\caption{$[[21,K,d_{min}]]$ Type-II QCS-A codes of different minimum distance $d_{min}$.}
\centering
\begin{tabular}{c c c}
\hline\hline
$[[N,K,d_{min}]]$ & $H_{proto}$ & Removed Rows \\ \hline
$[[21, 5, 5]]$ & $\left[H_{2proto}\boxplus\emptyset^{k\times k}|H_{1proto}\right]$ & \{7, 11, 12, 14, 21\}\\ \hline
$[[21, 5, 4]]$ & $\left[H_{1proto}|H_{2proto}\boxplus\emptyset^{k\times k}\right]$ & \{7, 11, 12, 14, 21\}\\ \hline
$[[21, 5, 3]]$ & $\left[H_{1proto}\boxplus\emptyset^{k\times k}|H_{2proto}\right]$ & \{5, 8, 9, 13, 21\}\\ \hline
$[[21, 6, 4]]$ & $\left[H_{1proto}\boxplus\emptyset^{k\times k}|H_{2proto}\right]$ & \{7, 11, 12, 14, 15, 21\} \\ \hline
$[[21, 6, 1]]$ & $\left[H_{2proto}\boxplus\emptyset^{k\times k}|H_{1proto}\right]$ & \{7, 11, 12, 14, 15, 21\} \\ \hline
$[[21, 6, 3]]$ & $\left[H_{2proto}\boxplus\emptyset^{k\times k}|H_{1proto}\right]$ & \{3, 7, 11, 12, 14,  21\} \\ \hline
$[[21, 6, 4]]$ & $\left[H_{1proto}|H_{2proto}\boxplus\emptyset^{k\times k}\right]$ & \{3, 7, 11, 12, 14, 21\}\\ \hline
$[[21, 6, 2]]$ & $\left[H_{2proto}\boxplus\emptyset^{k\times k}|H_{1proto}\right]$ & \{4, 7, 11, 12, 14,  21\} \\ \hline
$[[21, 6, 4]]$ & $\left[H_{1proto}|H_{2proto}\boxplus\emptyset^{k\times k}\right]$ & \{4, 7, 11, 12, 14, 21\}\\
\hline\hline
\end{tabular}
\label{tab:QCS-A21}
\end{table}

\section{Conclusion and Discussions}

In this work, two types of general quantum stabilizer codes were proposed based on quadratic residue sets of prime modulus. Different construction methods are proposed based on the property of quadratic residue sets and the constructed codes satisfy the commutative constraint.
The minimum distance for Type-I stabilizer codes of length $N=4n+1$ is closely related to the size of quadratic residue sets while the dimension of the codes is a constant. The code rate for Type-I stabilizer codes of length $N=4n-1$ is near half.
By exploiting the property between the elements of quadratic residue sets, we showed that each quadratic residue set is capable of designing proto-matrix of Latin Square format, and permitted a substantial number of new quasi-cyclic stabilizer codes constructed after lifting a pre-obtained proto-matrix.   
We demonstrated that the constructed new codes meet the distance bounds as shown in the literature.

\end{document}